\documentstyle[epsf,eqsecnum,aps]{revtex}
\begin{document}
\twocolumn

\wideabs{
\title{Bifurcations in annular electroconvection with an imposed shear}

\author{Zahir A. Daya$^{1,2}$, Vatche B. Deyirmenjian$^{1}$, and Stephen W.
Morris$^{1}$}
\address{$^{1}$Department of Physics, \\ University
of Toronto, Toronto, Ontario, Canada, M5S 1A7 \\
$^{2}$ Center for Nonlinear Studies,\\
Los Alamos National Laboratory, Los Alamos, NM 87545}
\date{\today}
\maketitle

\begin{abstract}

We report an experimental study of the primary bifurcation in
electrically-driven convection in a freely suspended film.  A weakly
conducting, submicron thick smectic liquid crystal film was supported by
concentric circular electrodes.  It electroconvected when a sufficiently
large voltage $V$ was applied between its inner and outer edges.  The film
could sustain rapid flows and yet remain strictly two-dimensional.  By
rotation of the inner electrode, a circular Couette shear could be
independently imposed. The control parameters were a dimensionless number
${\cal R}$, analogous to the Rayleigh number, which is $\propto V^2$ and
the Reynolds number ${\cal R}e$ of the azimuthal shear flow.  The
geometrical and material properties of the film were characterized by the
radius ratio $\alpha$, and a dimensionless number ${\cal P}$, analogous to
the Prandtl number. Using measurements of current-voltage characteristics
of a large number of films, we examined the onset of electroconvection
over a broad range of $\alpha$, ${\cal P}$ and ${\cal R}e$.  We compared
this data quantitatively to the results of linear stability theory.  This
could be done with essentially no adjustable parameters.  The
current-voltage data above onset were then used to infer the amplitude of
electroconvection in the weakly nonlinear regime by fitting them to a
steady-state amplitude equation of the Landau form. We show how the
primary bifurcation can be tuned between supercritical and subcritical by
changing $\alpha$ and ${\cal R}e$.

\end{abstract}
}

\section{Introduction}

Phenomological amplitude equation models are important tools in the study of
symmetry-breaking, pattern-forming bifurcations.\cite{ch93,NPL} They have
proven particularly useful at the weakly nonlinear level.  While their broad
application stems from the common symmetries that are shared by many different
systems, it is the coefficients or specific parameters in amplitude
equations that distinguish each system. It is an interesting question to
examine how these coefficients, especially those of the nonlinear terms,
change as the properties of the system are varied.  There are only a few
experimental systems, such as Rayleigh-B\'{e}nard convection, Taylor vortex
flow, and electrohydrodynamic convection in nematic liquid crystals, for which
these coefficients have been quantitatively determined.\cite{ch93}  

Three-dimensional, nonlinear systems are prone to develop complicated spatial and temporal
patterns even when only weakly nonequilibrium.\cite{ch93} The spatio-temporal patterns are often
the result of successive symmetry breaking bifurcations.  It is useful to study pattern
formation in low-dimensional systems which are close to equilibrium but have little symmetry so
that there are only a very limited set of symmetry-breaking bifurcations available.  In general,
one seeks the most complex dynamics that can be realized in as simple and restricted a system as
possible.  In our system, annular electroconvection, we exploit the strict two-dimensionality of
a submicron smectic A liquid crystal film.  The lower dimensionality greatly reduces the variety
of possible pattern states and so makes it easier to experimentally study the rich nonlinear
properties of the basic pattern.

In this paper, we report an experimental study of the properties of the bifurcations in
annular electroconvection with a variable circular Couette shear. Our primary focus is to
investigate the variation of the cubic nonlinear coefficient of an amplitude equation as the
experimental parameters are systematically changed.  This coefficient determines the
saturation behaviour of the convective flow velocity.  We begin by establishing that the
experimental system is well-described by a theoretical model that reduces to an amplitude
equation near the onset of the pattern-forming instability.  We test the adequacy of the
theoretical model by direct comparisons between the experimental results and theoretical
predictions at the linear level.\cite{DDM_pof_99}  We then proceed to describe how the 
coefficient of the
lowest order nonlinear term depends on the experimental parameters. Interestingly, we find
that the coefficient can be made to change sign and thus that the primary bifurcation can be
tuned between subcritical and supercritical.  In this paper, we show from symmetry
considerations that the amplitude equation that describes the sheared case is the complex
Ginzberg-Landau equation.\cite{ch93,NPL,knobloch} Elsewhere, we have derived this result 
directly from the full
electrohydrodynamic equations using a multiple-scales analysis.\cite{dey_annular_nonlinear} 
Calculation of the
parameter dependence of the coefficients of this equation is beyond the scope of this paper,
but in future, we expect to be able to compare the experimental results presented here with
theory at the weakly nonlinear level.

\begin{figure}
 \epsfxsize =3in
 \centerline{\epsffile{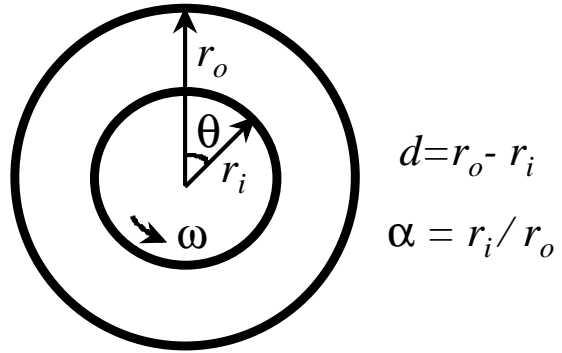}}
\vskip 0.2in 
\caption{The geometry of the annular film. Circular Couette shear is produced by rotating the
electrode holding the inner edge of the film.}
 \label{schematic} 
\end{figure}

Our system consists of a freely suspended annular smectic A film which can be driven out of
equilibrium by electrical forces and can be independently subjected to a shear flow as shown in
Fig.~\ref{schematic}. It was previously described in Refs.~\cite{DDM_pof_99}
and~\cite{annular98}. The film was suspended  between concentric circular electrodes and was
driven by a voltage difference between the electrodes.  The inner electrode could be rotated
about its axis, thereby imposing a circular Couette shear.  The flow pattern that emerges when
the film is driven sufficiently hard is referred to as electroconvection and consists of an
array of counter-rotating vortices.  Our primary probe of the electroconvection amplitude was
the excess electrical current carried convectively by the flow.  The annular geometry and
electrical nature of the experiment make it possible to study radially forced convection under
steady shear, a combination more often associated with geophysical flows than with small
laboratory systems. On the other hand, the accurately two-dimensional nature of the flow makes
it far simpler than most planetary analogs, and thus more amenable to theoretical investigation.

The driving force originates from the interaction of the radial electrical field and the surface
charge density that develops on the film's free surfaces.\cite{linear} Below the onset of
convection, the film's electrical state is independent of the imposed shear in the film. It is
thus straightforward to superpose shear flows with the radial driving, as these are
characterized by independent control parameters. The simplest such shear flow is steady
azimuthal or circular Couette flow.  The annular geometry is naturally periodic so that the
shear flow is closed and leads to a net mean flow around the annulus.

The addition of a shear flow to the annular electroconvection alters the symmetries of the base
state of electroconvection.  When shear is absent, the base state is invariant under azimuthal
rotation and reflection in any vertical plane containing the rotation axis through the center of
the annulus.  The electroconvective state is then stationary and appears with the spontaneous
breaking of the azimuthal invariance.\cite{annular98} When sheared, the reflection symmetry of the 
base state is not present. When driven, the pattern once again breaks the azimuthal symmetry, but since the
reflection symmetry is absent due to shear, the pattern is free to travel azimuthally in the
direction of the mean flow.\cite{annular98} In addition, as we show in detail below, the shear
flow alters the primary bifurcation, making it hysteretic.

The equilibrium properties of freely suspended liquid crystal films have been extensively
studied; see Ref.~\cite{sonin} for a recent review. Smectic liquid crystals consist of layers of
orientationally ordered long molecules which readily form suspended films.  In smectic A, the
average orientation of the long axis of the molecules, and hence the optic axis, is normal to 
the layer plane. The layers are of uniform thickness and
within each layer the distribution of molecules is isotropic.  Smectic A exhibits
two-dimensional isotropic fluid properties in the layer plane while flows perpendicular to the
layers are strongly inhibited.  Other material properties such as the electrical conductivity
and dielectric permitivity are also isotropic in the plane of the layers.  Uniform suspended
smectic films are always an integer number of smectic layers thick and while they flow readily,
they seldom change thickness; 

\noindent they are robustly two-dimensional.  Unlike soap
solutions\cite{airdrag,couder}, smectic liquid crystals have very small vapor pressures, hence
the film can be enclosed in an evacuated environment.  The reduction in ambient pressure leads
to a proportional reduction in the air drag to which the film is subjected.\cite{airdrag} 
This system thus has several attractive features for the study of bifurcations in simple nonlinear
systems that may be described by amplitude equation models. Our study greatly extends and complements 
previous work on electroconvection in isotropic liquids\cite{avsec,malkus,jolly,melcher,felici,agrait88,castellanos}, 
nematics\cite{faetti,ecrev}, smectic A\cite{DDM_pof_99,annular98,linear,smorris1,smorris2,smorris3,smao_stuff,endselection,gle97} and higher smectic 
phases\cite{smctheory,smcexpt,langer,langer_annular}.

The paper is organized as follows. In Section~\ref{lineartheory} we describe the results of a
linear stability analysis of our system. In the process we define the various dimensionless
parameters that characterize the experiment. In Section~\ref{amplitudeequation}, we use symmetry
arguments to justify the form of the relevant amplitude equation.  Our description of the
experimental apparatus and protocol follows in Section~\ref{experiment}.  In
Section~\ref{dataanalysis} we outline our data analysis procedure.  We present experimental
results that pertain to linear analysis in Section~\ref{comparisonlinear}. An experimental
determination of the cubic coefficient of the amplitude equation is found in
Sections~\ref{gRe=0} and~\ref{gRe}.  Section~\ref{discussion} discusses the relationships
between our results and those for other similar systems and section~\ref{conclusion} presents a
brief summary and conclusion.

\section{Theoretical Background}

In this section, we briefly review the linear and weakly nonlinear theory
that is relevant to our data analysis and results.

A linear stability analysis of annular electroconvection with circular Couette flow was reported in
Ref.~\cite{DDM_pof_99}. The theory is constructed with a number of simplifying
assumptions.\cite{DDM_pof_99,linear} The fluid film is treated as an annular sheet of inner radius $r_i$,
outer radius $r_o$ and thickness $s$.  A schematic is shown in
Fig.~\ref{schematic}. An important geometric parameter is the radius ratio $\alpha = r_i/r_o$.
The film width $d = r_o - r_i$ is assumed to be much greater than $s$. The viscous fluid is
assumed to flow only in two-dimensions, and be incompressible, Ohmic and of uniform electrical
conductivity. We denote the fluid density by $\rho$, its molecular viscosity by $\eta$ and its
electrical conductivity by $\sigma$.  Only the charges due to free surfaces are included and
bulk dielectric effects inside the film are neglected.\cite{linear} The electrodes are assumed
to be of negligible thickness, and fill the rest of the plane not occupied by the annular
film. Outside the plane of the film, we assume an empty space of dielectric permittivity
$\epsilon_0$.  Couette shear in the theoretical model is imposed by specifying the appropriate
velocity boundary conditions on the edges of the film. Linear theory predicts the position of
the onset of convection and the degree to which onset is suppressed by the shear.  These are
discussed in detail in Ref.~\cite{DDM_pof_99} and are further analyzed in
Section~\ref{comparisonlinear} below.

From very general symmetry considerations, we can deduce the form of an
amplitude equation which is valid in the weakly nonlinear regime just above
onset.\cite{ch93,NPL,knobloch}  Here, a complex-valued amplitude describes
the slowly-varying magnitude and phase of the spatially periodic physical
fields, for example the fluid velocity.  We find an expression of the
well-known Landau form.\cite{dey_annular_nonlinear}  The coefficients of the
amplitude equation can in principle be calculated from the underlying
electrohydrodynamic equations; a calculation of this kind is in progress and
will be reported separately\cite{dey_annular_nonlinear}.  The magnitude of
the complex amplitude is directly related to the quantity we measured
experimentally, the total current carried by the film. We can thus fit
current-voltage data to the real part of the amplitude equation to
experimentally determine its coefficients. This is done in Sections~\ref{gRe=0}
and~\ref{gRe} below.

\subsection{Outline of linear stability theory}
\label{lineartheory}

The theoretical treatment of Ref.~\cite{DDM_pof_99} examined the linear stability of the
two-dimensional annular fluid film when it is subjected to a voltage $V$ at the inner electrode
of the annulus while the outer electrode is held at ground potential. In addition, the theory
allowed for the rotation of the inner edge of the annulus at angular frequency $\omega$, while
the outer edge is held fixed.  The rotation drives a Couette shear flow in the film below the
onset of convection.  It can be shown that any arbitrary rotation of the inner and outer edges
can be transformed to a rotation of only the inner edge by moving to the appropriate rotating
coordinates.\cite{DDM_pof_99,alonso95} Such a coordinate transformation has no effect on the
dynamics as long as the flow is strictly two-dimensional. The theoretical model is completely
specified by the radius ratio $\alpha$ and three additional dimensionless parameters:
\begin{equation} 
{\cal R} \equiv \frac{\epsilon_0^2 V^2}{\sigma \eta s^2} \,, \hspace{5mm} {\cal
P} \equiv \frac{\epsilon_0 \eta}{\rho \sigma s d} \,, \hspace{5mm} {\rm and} \hspace{5mm} {\cal
R}e \equiv \frac{\rho \omega r_i d}{\eta} \,. \label{Rayleigh} \end{equation}
 The control
parameter ${\cal R}$ is a measure of the external driving force which is proportional to the
square of the applied voltage $V$ while ${\cal P}$ is a ratio of the time scales of electrical
and viscous dissipation processes in the film.  The Reynolds number ${\cal R}e$ is a measure of
the strength of the applied shear, which is regarded as a second control parameter.  It has been
established\cite{DDM_pof_99,annular98} that the instability leads to a one-dimensional pattern
of $m$ vortex pairs where $m$ is the azimuthal mode number.  Linear stability predicts the value
of the critical control parameter ${\cal R}_c$ and the critical mode number $m_c$ at which the
film becomes marginally unstable.  ${\cal R}_c$ and $m_c$ are in general functions of $\alpha$,
${\cal P}$, and ${\cal R}e$. A detailed discussion of the calculated values of ${\cal R}_c$ and
$m_c$ under various conditions is given in Ref.~\cite{DDM_pof_99}.  When ${\cal R}e = 0$ ({\it
i.e.} with no applied shear), it was found that ${\cal R}_c$ and $m_c$ are independent of ${\cal
P}$. In the following, these critical values for zero shear will be denoted ${\cal R}_c^0$ and
$m_c^0$.

When ${\cal R}e > 0$, it was found that ${\cal R}_c > {{\cal R}_c}^0$, for
any $\alpha$ and ${\cal P}$, {\it i.e.} that the shear always suppresses the
onset of convection.\cite{mc_note}  It is convenient to measure the relative
degree of suppression for a given $\alpha$ and ${\cal P}$ in terms of the
reduced quantity
\begin{equation}
\tilde{\epsilon}~(\alpha,{{\cal R}e},{\cal P}) = \Biggl[\frac{{\cal
R}_c(\alpha,{{\cal R}e},{\cal P})}{{\cal R}_c^0(\alpha)}\Biggr] -1 \,,
\label{eptilde}
\end{equation}
which is a monotonically increasing function
of ${\cal R}e$ for all ${\cal P}$. In Section~\ref{comparisonlinear} below, we
extend the
discussion in Ref~\cite{DDM_pof_99} by comparing these results to data for
several different $\alpha$, and for a wide range of ${\cal P}$.

\subsection{Nonlinear theory: The amplitude equation}
\label{amplitudeequation}
The amplitude equation appropriate to our system follows from symmetry
considerations.\cite{ch93,NPL,knobloch}  We defer a detailed calculation of
its numerical coefficients to future work.\cite{dey_annular_nonlinear}

In the absence of shear, the base state solution of the electrohydrodynamic
equations is invariant under azimuthal rotations and reflections through any
plane perpendicular to the annular plane and containing its center. Just above onset, we assume
that all the
physical fields are nonaxisymmetric and proportional to a rapidly-varying spatial oscillation
$e^{ i m \theta}$,
where $m$ is an   
azimuthal mode number. This
requires
that the differential equation for the slowly-varying complex amplitude $A_m$ of mode $m$, be unchanged under
the transformations
\begin{equation}
\theta \rightarrow \theta + \theta' \,, \;\;\; A_m \rightarrow A_me^{im\theta'}
\;, \label{rotation}
\end{equation}
and
\begin{equation}
\theta \rightarrow -\theta  \,, \;\;\; A_m \rightarrow \bar{A}_m \;.
\label{reflection}
\end{equation}
In the above, $\theta$ is an azimuthal angle, and the overbar denotes
complex conjugation. The most general amplitude 
\begin{figure}
\epsfxsize =3.2in
\centerline{\epsffile{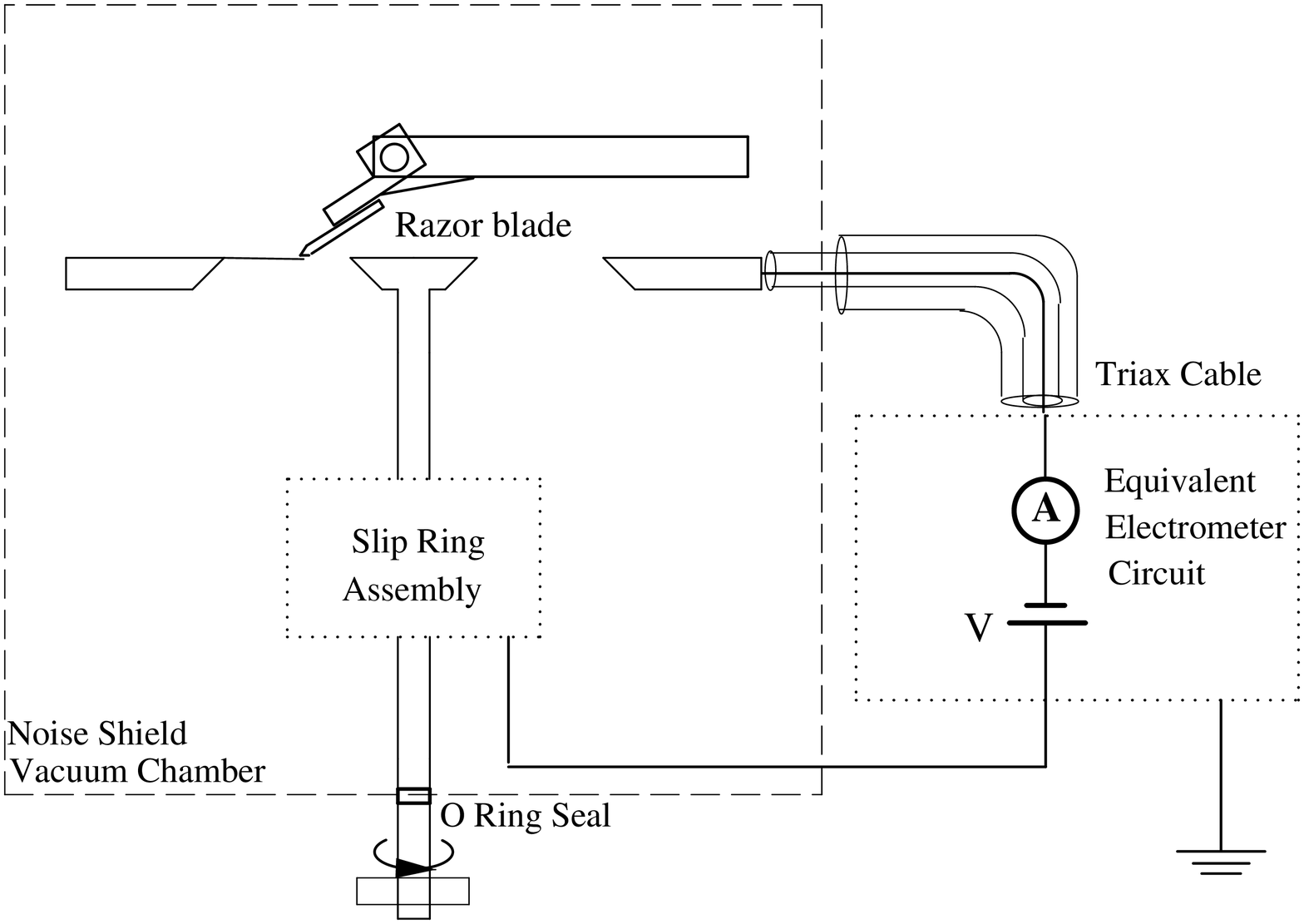}}
\vskip 0.2in
\caption{Schematic of the electrical and mechanical parts of the apparatus, showing 
the annular electrodes and film drawing assembly in side view.}
\label{schematic_expt}
\end{figure}
\vskip 0.2in

\noindent equation that is invariant
under these symmetry operations has the Landau form,
\begin{eqnarray}
\tau \partial_t A_m & = & \epsilon A_m - g |A_m|^2 A_m + h |A_m|^4 A_m
 - ... \;,
\label{landau}
\end{eqnarray}
where $\tau$, $g$, and $h$ are real-valued coefficients. The small
parameter $\epsilon$ is a reduced control parameter given by
\begin{equation}
{\epsilon} = \Biggl[\frac{{\cal R}}{{\cal R}_c}\Biggr]-1 \,,
\label{ep}
\end{equation}
and is a measure of the distance from threshold. The amplitude equation
is accurate for small $\epsilon$ and describes a bifurcation from the
$A_m \equiv 0$ state ($\epsilon < 0$) to the $A_m \neq 0$ state with $2m$
vortices
($\epsilon > 0$). A sharp bifurcation occurs at $\epsilon = 0$.

If the fluid is subjected to a circular Couette shear, the base state is
only invariant under azimuthal rotations, Eqn.~\ref{rotation}. In this case,
the general amplitude equation takes the {\it complex} Landau form
\begin{eqnarray}
\tau (\partial_t - ia_{Im}) A_m & = & \epsilon (1+ic_0) A_m -
g (1+ic_2) |A_m|^2 A_m \nonumber \\
 & + & h (1+ic_3) |A_m|^4 A_m
 - ... \;, \label{complexlandau}
\end{eqnarray}
where $a_{Im}$ is the imaginary part of the eigenvalue of the unstable mode
$m$ at onset and the coefficients $c_0$, $c_2$, $c_3$ are real-valued.
Eqn.~\ref{complexlandau}, which here is simply deduced from symmetry
considerations, has been rigorously derived for annular electroconvection
with shear from the basic electrohydrodynamic
equations.\cite{dey_annular_nonlinear} In general, it describes a Hopf
bifurcation to a pattern with a complex amplitude $A_m$ which travels in one
azimuthal direction.

In both Eqns.~\ref{landau} and \ref{complexlandau}, we have omitted terms
involving the azimuthal gradients of $A_m$, which if included would give the
so-called Ginzburg-Landau equation.\cite{ch93}  While such terms are
potentially quite interesting, they do not appear to be directly relevant to
interpreting our current-voltage data.  Our annular system is azimuthally
periodic and thus lacks lateral boundaries at which $A_m \rightarrow 0$ where
gradients would be important.\cite{endselection}  Thus, any gradients in $A_m$
would have to arise from purely dynamical effects, such as Eckhaus or other
phase instabilities\cite{ch93}.

From current-voltage data, we can extract the reduced Nusselt number $n$,
which is a global dimensionless quantity defined as the ratio of
current carried by convection to that by conduction. The reduced Nusselt number
$n$ is related to the electric Nusselt number ${\cal N}$ by $n = {\cal N}-1$.
One can show\cite{gle97,dey_annular_nonlinear} that the amplitude can
be scaled so that $n = |A_m|^2$.

To solve for the real and imaginary parts of the amplitude equation, let
$A_m(t)=A(t)e^{i\Phi(t)}$, where $A(t)$ and $\Phi(t)$ are real. Substitution
into Eqn.~\ref{complexlandau}, gives
\begin{eqnarray}
\tau \partial_t A & = & \epsilon A - g A^3
 - h A^5 - ... \;, \label{AMP6_real} \\
\tau(\partial_t \Phi - a_{Im}) & = & \epsilon c_0 -g c_2 A^2 + ...
\;. \label{AMP6_imag}
\end{eqnarray}
We will show in Section~\ref{dataanalysis} how the raw current-voltage data
can be transformed into measurements of $\epsilon$ and $n$, and hence
be used to measure the real amplitude $A$ via a nonlinear fit procedure.  We
focus on
extracting the coefficient $g$ of the cubic nonlinearity. The sign of $g$
determines
whether the pitchfork bifurcation to electroconvection is supercritical
(``forward'', $ g > 0$ )
or subcritical (``backward'', $ g < 0$ ).  A tricritical bifurcation occurs
when $g = 0$.
The results for $g$ are discussed in Sections~\ref{gRe=0} and~\ref{gRe}.

\section{Experimental Apparatus}
\label{experiment}
In this Section we describe the liquid
crystal sample and the experimental apparatus.  We performed a large number of
precise current-voltage measurements on electroconvecting annular films with a
wide range of radius ratios and applied shears.  In overall structure, the
apparatus consisted of the annular electrodes which were housed in a vacuum
chamber, an electrometer circuit and a stepper motor, both under computer
control. The chamber also allowed films to be drawn under vacuum and
inspected under reflected white light to check thickness uniformity. A
schematic is shown in Fig.~\ref{schematic_expt}.

\subsection{The liquid crystal}
As in previous experiments\cite{DDM_pof_99,annular98,smorris1,smorris2,smorris3,smao_stuff,endselection}, our study used
smectic A octylcyanobiphenyl (8CB). 8CB has a smectic A phase
between ${\rm 21^{\circ}C}$ and ${\rm 33.5^{\circ}C}$.  The electrical
conductivity of ``pure'' 8CB is due to several ionic impurities of
varied and unknown concentrations.   In order to control the nature of the
ionic species contributing to the electrical conductivity, we doped the 8CB
with tetracyanoquinodimethane (TCNQ), a material believed to form charge
transfer complexes with the host, so that the dominant impurity species was
the dopant.  To prepare the doped material, TCNQ was dissolved in
acetonitrile and added to the 8CB sample.  The acetonitrile was then
evaporated in a vacuum oven while warming the mixture so that the 8CB was in
its isotropic phase.  The samples used had concentrations of
$2.96 \times 10^{-4}$, $1.11 \times 10^{-4}$ and $7.62 \times 10^{-5}$, of
TCNQ by weight. Experiments with significantly higher or lower dopant
concentrations were found to be less useful due to irreproducible non-ohmic
behaviour below the onset of convection.\cite{dope_note}

\subsection{the annulus}
The annular electrodes were constructed out of stainless steel.
The inner electrode was a circular disk of diameter $2r_i$.  The outer
electrode was a circular plate of diameter $9.00$~cm  with a central hole of
diameter $2r_o$.  The outer electrode was $0.73 \pm 0.01$~mm thick.  By using
several pairs of inner and outer electrodes, we conducted experiments at six
different radius ratios between $\alpha = 0.33$ and $\alpha = 0.80$.  We
used inner electrodes with radii ranging between $r_i = 3.60 \pm 0.01$~mm and
$r_i = 5.26 \pm 0.01$~mm.  The radii of the outer electrodes were between
$r_o = 5.57 \pm 0.01$~mm and $r_o = 11.25 \pm 0.01$~mm.

The annulus was housed in a vacuum chamber which was evacuated by a rotary
pump.  The air was evacuated slowly so as to prevent vigorous air flows that
may cause the film to rupture.  The surrounding air was pumped down to an
ambient pressure between $0.1-5.0$~torr.  At these pressures, the mean free
paths of $N_2$ and $O_2$ are in the range $0.5-0.05$~mm. This was comparable
to the film width $d$, so that the air drag on the film
was negligible.\cite{airdrag}   All experiments were performed at the ambient
room temperature of $23 \pm 1^\circ$C, well below the smectic A-nematic
transition at $33.5^\circ$C for undoped 8CB.

The inner electrode was made to rotate about its axis by means of a high
precision stepper motor operating at $25600$~steps per revolution. The motor
was located outside the vacuum chamber and was connected through a rotating
seal.  The inner electrode was adjusted to rotate true to its axis to within
$50\mu$m at angular frequencies up to $\omega = 6\pi$~rad/s.

\subsection{Drawing the film and determining its thickness}
A motorized film-drawing assembly was housed in the vacuum chamber. The film
was spread across the annular gap between the electrodes using a stainless
steel razor blade inclined at $\sim 25^\circ$.  The blade was moved away from
the annulus while voltage was applied, to avoid perturbing the electric
fields near the film.

The thickness of the film was determined from the interference color of the
film under reflected white light using a low power videomicroscope.   Using
standard colorimetric functions\cite{colorimetry,colordynamics,sirota} a
color-thickness chart was calculated for 8CB.  Since smectic films are
constrained to be an integer number of smectic layers thick, the film color
can be used to identify the film thickness measured in smectic layers.  Each
layer of smectic A 8CB is $3.16$~nm thick.\cite{thicknessSmA}  Most of the
experiments were performed with films between $25$ and $85$ layers thick.
Over most of the middle of this range, the film thickness can be determined
to within $\pm 2$ layers, while close to the ends of the range a more
conservative estimate of $\pm 5$ layers was used.  The
colorimetric method worked well for film thicknesses where the strong
interference color can be unambiguously matched to a color chart.  Very thin
films that appeared black and thicker, nearly white films were not used.

During the course of an experiment, the film color was monitored to check
that it remained uniform in thickness to within $\pm 1$ layer.  Films
drawn with non-uniform thicknesses, left to themselves, tend to anneal to
become uniform in thickness and hence uniform in color. The annealing process
can be accelerated by electroconvecting and shearing the films.

Current-voltage runs during which the film thickness spontaneously became
non-uniform were abandoned. A small number of visualization experiments were
performed on films which had a nonuniformity in thickness of about $\pm
2$~smectic layers, or about $5\%$ of $s$. Usually these nonuniform films had
two thicknesses and in reflection displayed two nearby interference colors. The
advection of the thickness nonuniformities was used to visualize the flow
pattern and thus to determine the azimuthal mode number $m$ by counting
vortices. For simplicity, some of these observations were done at atmospheric
pressure. This visualization method, while somewhat crude, was preferrable to
using suspended smoke particles, which tend to perturb the conductivity of the
film.\cite{smao_stuff}

\subsection{Current-voltage measurements}
Since the current transported through the film was picoamperes in magnitude,
particular care had to be exercised to avoid stray currents. The inner
rotating electrode was electrically isolated from all other components by a
highly insulating sleeve. The current was carried onto the rotating electrode
by a set of silver-graphite slip rings inside the vacuum chamber.  A Keithley
model 6517 electrometer was used both as a voltage source and a picoammeter.
The `high' of the programmable dc voltage source of the electrometer was
connected to the rotating inner electrode, which is the only part of the entire
apparatus that is not at ground potential.  The outer annulus was connected to the input of 
the electrometer which is held at very nearly ground potential. The outer annulus was 
electrically isolated from true ground by teflon washers which served to eliminate leakage 
currents which would otherwise be added to the signal.   The rest
of the apparatus was grounded. Electrical noise was
reduced by shielding the electrodes and most of the 
\begin{figure}
\epsfxsize =3in
        \centerline{\epsffile{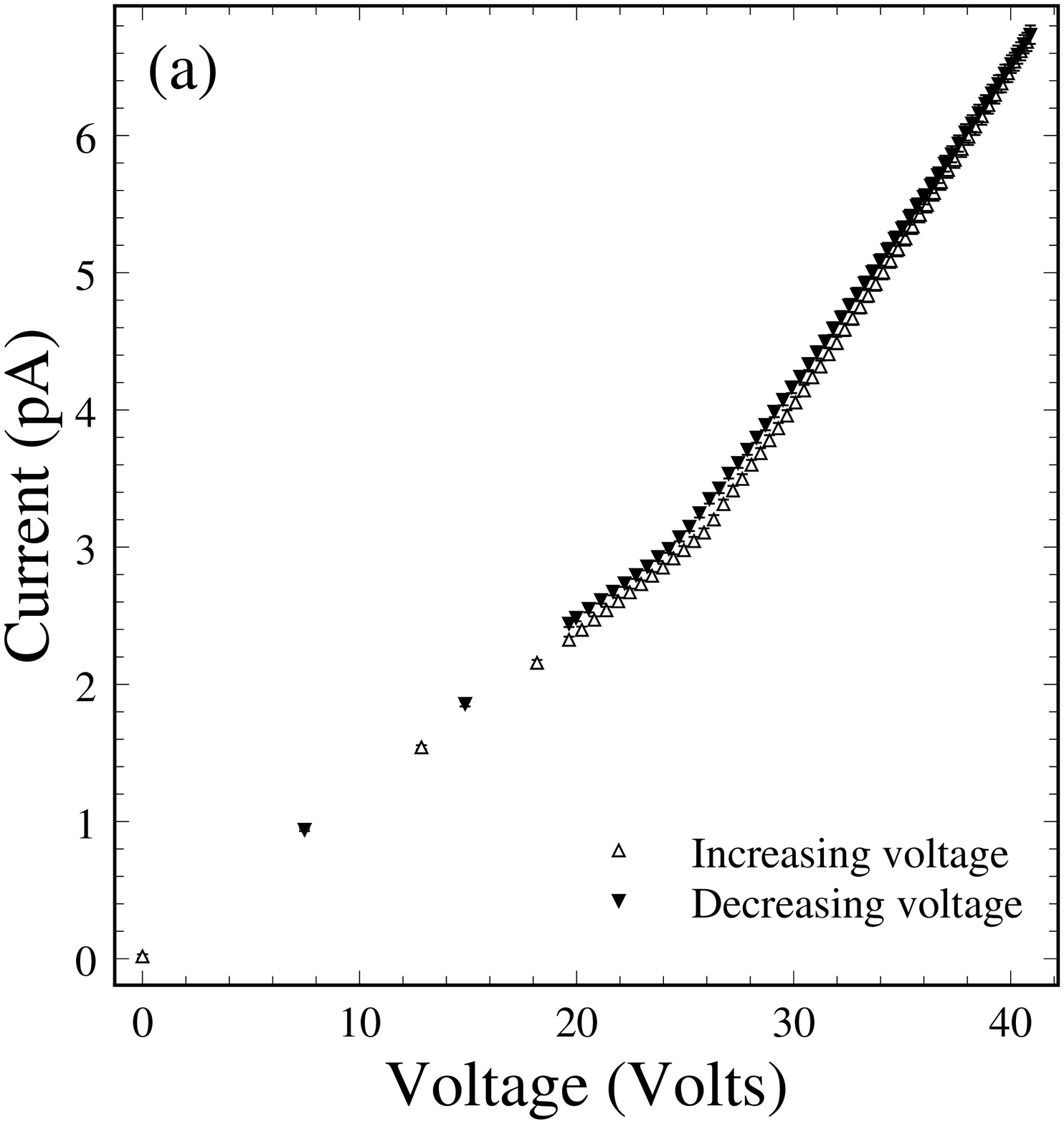}}
        \vskip 0.1in
\epsfxsize =3in
        \centerline{\epsffile{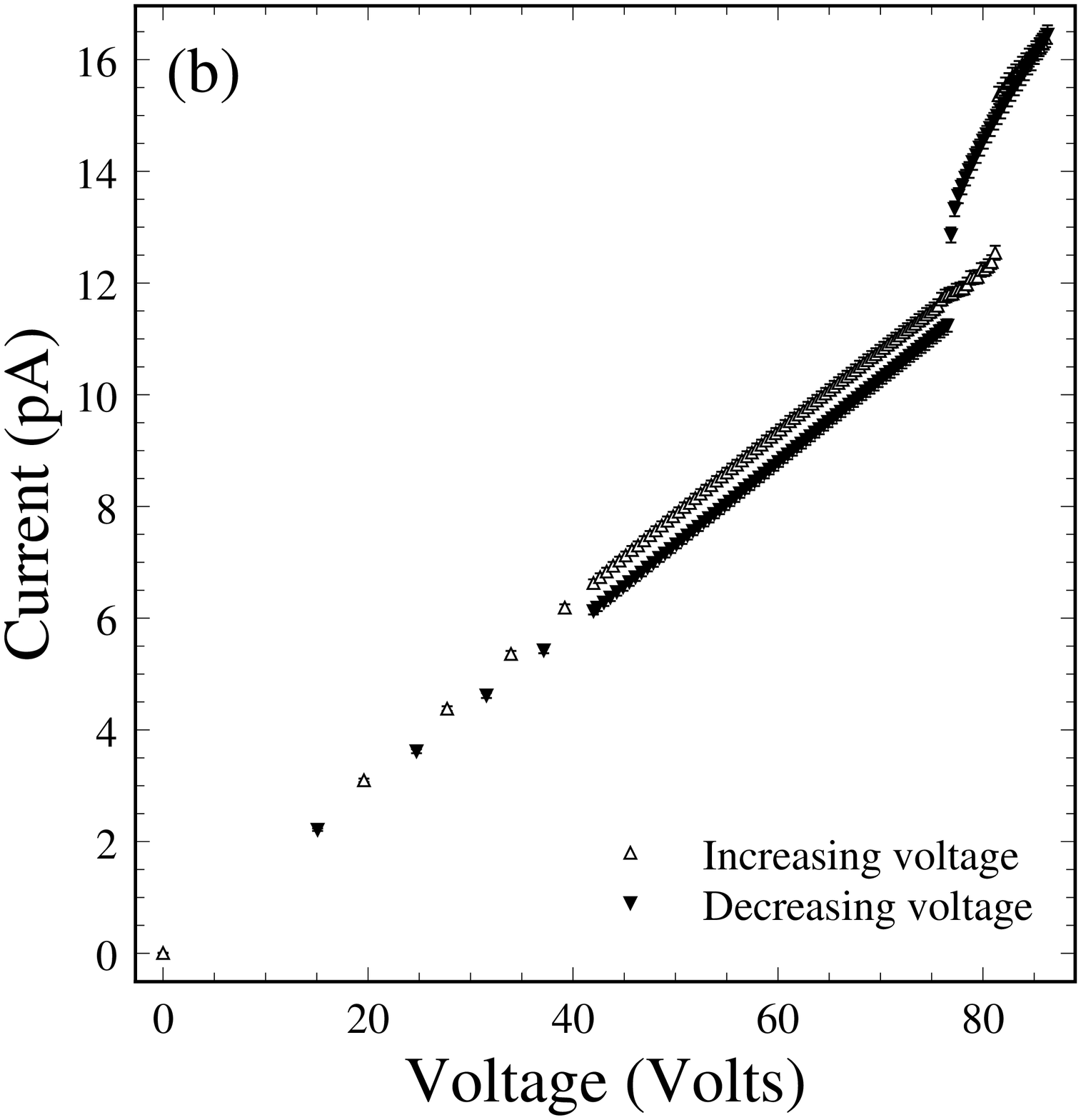}}
        \vskip 0.1in
        \caption{Representative
current-voltage characteristics for radius ratio $\alpha = 0.467$ in the absence of shear
(a) and when strongly sheared (b).  Note the different scales.}
        \label{experiment_IV}
\end{figure}

\noindent experimental components
in a large Faraday cage which doubled as the vacuum chamber.  Low noise
triaxial cables and feedthroughs were used to connect the outer electrode to
the electrometer.

An example of a current-voltage characteristic in a film without shear is
shown in Fig.~\ref{experiment_IV}a. It consists of data obtained for
incremental and decremental voltages. The current-voltage characteristic
clearly shows two regions: one for voltages smaller than a critical voltage
$V_c$ and one for voltages greater than $V_c$.  The critical voltage that
separates these two regions is in the vicinity of the kink in the
current-voltage characteristic.  In the regime $V \leq V_c$, the current is
linearly dependent on the voltage and the film is ohmic.
Experiments in 8CB with significantly higher and lower concentrations of TCNQ
show, at least initially, non-ohmic current-voltage characteristics even for
$V \leq V_c$, due to electrochemical effects.

We estimated $V_c$ before each run then defined an approximate reduced
control parameter $\epsilon= (V/{V_c}^{est})^2 - 1$.  Current-voltage data
was then obtained by making variable steps in voltage in such a way as to
make equal increments in $\epsilon$.
We used several $\epsilon$ step sizes, which became finer near
threshold.  Similarly, we used a variable waiting time after each step, which
allowed for longer relaxation times closer to $V_c$. Very long relaxation
times were not feasible due to the drift of the electrical
conductivity.\cite{drift}
After the wait time, between $100 - 200$
measurements of the current, each separated by $25$~milliseconds, were
averaged.  The error in the average was taken to be the standard deviation of
the mean or $1\%$, the reading error of the picoammeter, whichever was
greater.  Each run consisited of incrementing $\epsilon$ up to a predetermined
maximum
and then decrementing it to zero voltage again.  Once the data had been
acquired, more
precise $\epsilon$ values at each voltage were found using drift-corrected
values of
$V_c$ from a fit procedure described in the next section.

When the film was sheared by rotation of the inner electrode, it was allowed
at least $30$~seconds after a change of shear rate to attain a steady state
before the current-voltage characteristics were obtained.  In all runs, the
shear
rate was established and then held fixed during a current-voltage sweep.
Fig.~\ref{experiment_IV}b displays a representative current-voltage
characteristic in a film under shear.  The principle effects of the shear are
to suppress the onset of convection and, eventually, to make the primary
bifurcation hysteretic.

\section{Experimental Results}
\label{results}

With the exception of some simple flow visualization which used slightly
nonuniform films, all our results were obtained by fitting current-voltage data
taken with uniform films.  It is possible to fit the data in such way as to
extract all the unknown material parameters in a nearly model-independent way,
as well as to correct for small drifts in conductivity.

The first task was to determine the dimensionless parameters relevant to each
current-voltage sweep: the radius ratio $\alpha$, the dimensionless ratio 
${\cal P}$ and the Reynolds number ${\cal R}e$. We can then properly
non-dimensionalize the current-voltage data and fit it to an amplitude equation
model.  The details of this fit are described in the section \ref{dataanalysis}
below.

In Section~\ref{comparisonlinear}, we consider the features of the data that
are predicted by the linear stability theory developed in
Ref.~\cite{DDM_pof_99}. For zero shear, these are the critical mode number
$m_c^0$, and the critical voltage $V_c^0$.  We use the latter to fix the last
unknown material parameter, the viscosity $\eta$.  We also compare our data to
the linear stability prediction for the shear-suppression of the onset of
convection. This was done for six values of $\alpha$, and for wide ranges of
${\cal P}$ and ${\cal R}e$.

In the weakly nonlinear regime, the value of the coefficient of the cubic
nonlinearity, $g$ in the amplitude equation, was of primary interest.  In
Section~\ref{gRe=0} we report results for $g$ at various $\alpha$ and ${\cal
P}$ for ${\cal R}e = 0$, that is, in the absence of shear.  We find that $g$
can become slightly negative for small $\alpha$, so that the bifurcation is
weakly backward.  The result for $g$ in the limit ${\cal R}e = 0$, $\alpha
\rightarrow 1$ is compared to the theoretical result for rectangular
films\cite{gle97}. Finally, in Section~\ref{gRe} we describe the results for
$g$ in sheared films, where ${\cal R}e > 0$.   We find that increasing ${\cal
R}e$ has a strong effect on the nature of the bifurcation, driving $g$ negative
and making the bifurcation backward.

\subsection{Fitting the current-voltage data}
\label{dataanalysis}

Except for a small drift, the film is ohmic below the onset of convection.  We
used fits in this range to determine some important material parameters in
order to scale the data.

Each voltage-current measurement in the ohmic regime,
$(V,I)$, constitutes an experimental determination of the film's
conductance, $c = I/V$.  For
a film of radius ratio $\alpha$, thickness $s$ and
conductivity $\sigma$, the conductance is given by

\begin{equation}
c = \frac{2\pi \sigma s}{\ln( 1/\alpha )} \,.
\label{conductance}
\end{equation}
Interestingly, the conductance is independent of the size of the film, {\em
i.e.} independent of $r_i$ or $r_o$. Since $\alpha$ is merely a geometrical
parameter, measurements of $c$ are effectively measurements of $\sigma s$,
eliminating the need to determine these separately.

Using Eqn.~\ref{conductance}, ${\cal P}$, defined in
Eqn.~\ref{Rayleigh}, can be expressed in terms of $c$ as

\begin{equation}
\label{pdimensional} {\cal P} = \frac{\epsilon_0
\eta}{\rho \sigma
s d} = \frac{2\pi \epsilon_0 \eta}{\rho
(r_o - r_i) \ln{( 1/\alpha )}}~~\frac{1}{c}\,,
\end{equation}
where $d = r_o - r_i$.  Similarly, the Reynolds number ${\cal R}e = \rho \omega
r_i d / \eta$ can be calculated from the measured angular frequency $\omega$ of
the inner electrode in rad/s, given $\rho$ and $\eta$.   The density $\rho$ of
8CB at room temperature\cite{8CBdensity} is $1.0 \times 10^{3}$ kg/m$^3$.  In
both ${\cal P}$ and ${\cal R}e$ , the only remaining undetermined material
parameter is the viscosity $\eta$. We fixed this parameter using a combination
of experimental and theoretical results, as described in
Section~\ref{comparisonlinear}. The result was $\eta = 0.18 \pm 0.03$ kg/ms.

Thus, for each current-voltage sweep, we were able to determine ${\cal P}$ and
${\cal R}e$.  The drift in conductivity\cite{drift} 
caused a drift in $c$ and hence in ${\cal P}$ over the course of a sweep.
Furthermore, $c$ is only known directly during the ohmic parts at the beginning
and end of each sweep.  We associated a mean ${\cal P}$ with each sweep by
averaging over $c$ data before and after.  This results in $\sim 10\%$ errors
in ${\cal P}$ for any one sweep, and a slow, uncontrolled evolution of ${\cal
P}$ from sweep to sweep.

The whole range of current-voltage data is fit to the amplitude equation model.
 Eqn.~\ref{complexlandau} gives the full, time-dependent complex coefficient
amplitude
equation required by symmetry.  Only the real and steady
state part of Eqn.~\ref{complexlandau}, given by

\begin{equation}
\label{amplitudemodel} \epsilon A - g A^3 - h A^5 + f = 0\,,
\end{equation}
is required to model the current-voltage data.  Here we have truncated at the
quintic order and augmented Eqn.~\ref{complexlandau} with a phenomenological
field term $f$.  This term models the rounding of the bifurcation due to
nonideal systematic effects which are
slightly symmetry-breaking, resulting in an ``imperfect''
bifurcation.\cite{ch93}  For all of the fits, we found $f \ll 1$.

As discussed in section \ref{amplitudeequation}, the amplitude $A$ which
appears in Eqn.~\ref{amplitudemodel} is related to the reduced Nusselt number
$n$ by
\begin{equation}
\label{n} n = \frac{I}{I_{cond}} - 1 = \frac{I}{c V} - 1 = A^2 \,,
\end{equation}
while the reduced control parameter $\epsilon$ is given by
\begin{equation}
\label{epsilon} \epsilon = \frac{{\cal R}}{{\cal R}_c} - 1 =
\Biggl(\frac{V}{V_c}\Biggr)^2 - 1\,.
\end{equation}
Eqns~\ref{amplitudemodel}-\ref{epsilon} contain five parameters, $c$, $V_c$,
$g$, $h$, and $f$.  While $c$ has a nearly constant initial slope and $V_c$ is
marked by a relatively obvious kink in the raw $(I,V)$ data, the amplitude $A$
is indirectly deduced via the pair of transformations Eqns.~\ref{epsilon} and
\ref{n} which are nonlinear.  This amplitude
is in turn fit using Eqn.~\ref{amplitudemodel}, which is again
nonlinear.  Thus, the parameters $(g,h,f)$ are rather distantly related to
the raw $(I,V)$ data.  If $g<0$, the bifurcation is hysteretic, and $A$ is
multivalued over some ranges of $\epsilon$.  Also, the nature of the model
necessarily involves
several fit parameters which are not independent.  Consequently, the
determination of these parameters is much more difficult than $V_c$.  As
discussed in Sections~\ref{gRe=0} and \ref{gRe},  $(g,h,f)$ can be influenced
by small systematic effects in the data leading to scatter which is larger
than the statistical uncertainties in the fits.  Nevertheless, the general
trends are robust.

The drift of the film's conductivity, and hence its conductance $c$, introduces
some additional complications.  Recall that the onset of convection occurs when
the control parameter ${\cal R}$
equals a critical value ${\cal R}_c$ given by
\begin{equation}
{\cal R}_c =\frac{\epsilon_0^2 {V_c}^2}{\sigma \eta s^2}~{\rm or}~V_c
=\frac{s}{\epsilon_0} \sqrt{{\cal R}_c \sigma \eta}\,. \label{RcVc}
\end{equation}
Since the electrical conductivity is not constant, the value of $V_c$ required
to calculate $\epsilon$ slowly changes during the course of an experiment.
Combining Eqns.~\ref{conductance} and \ref{RcVc}
yields
\begin{equation}
\label{Vcc}
V_c(c) = \sqrt{\frac{s\eta{\cal R}_c~{\ln( 1/\alpha )}~c }{2\pi
{\epsilon_0}^2}}\,.
\end{equation}
\begin{figure}
\epsfxsize =3in
\centerline{\epsffile{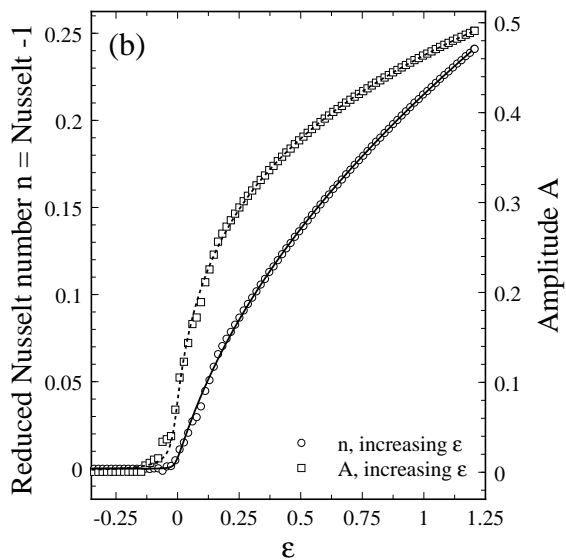}}
\vskip 0.15in
\caption{The amplitude 
$A$ and the reduced Nusselt number $n=A^2$ 
{\it vs.} the control parameter
$\epsilon$ for a film with radius
ratio $\alpha = 0.64$.  The solid and dashed lines are 
a fit to the Landau amplitude equation.}
\label{supercritical_plot}
\end{figure}
\vskip 0.2in

\noindent We corrected for the drift in $c$ by tracking its value during the ohmic parts
of the sweep and using a linear interpolation during the portion of the sweep
where the film is convecting.  In this way, {\it each} voltage and current
measurement was non-dimensionalized with its own value of $c$ and $V_c(c)$ to
produce fittable values of $\epsilon$ and $A$.  We also transformed the errors
in the current measurements $\Delta I$ into amplitude errors $\Delta A$.

We approached the fitting problem in the following bootstrap fashion.  It was
easy to
ascertain bounds on $V_c$ by inspecting the current-voltage characteristics, see for
example Figs.~\ref{experiment_IV}a and b.  Each current-voltage
characteristic was scrutinized and two voltage intervals were chosen.
The first interval contained the critical voltage at which the
conducting state became unstable to the convecting state.
The second interval contained the voltage at which the convecting
state became marginally unstable to the conduction state.
Guesses of $V_c$ in both these intervals
were chosen at random using a uniform deviate random number
generator\cite{recipes}.  The conductances $c$ were then determined for the
ohmic regimes and from a linear interpolation for the convecting portion, as
described earlier.  The raw current-voltage data was then transformed into
$\epsilon$ and $A \pm \Delta A$ and fit to Eqn.~\ref{amplitudemodel} by varying
only the three parameters $g$, $h$ and $f$ in a weighted Levenberg-Marquardt
nonlinear fitting procedure which minimized $\chi^2$.  We then did a
Monte-Carlo optimization of the randomly chosen
$V_c$.\cite{recipes,constraint}.  The best $V_c$ was taken to be the one that
minimized $\chi^2$ over the Monte Carlo sample.  The uncertainty in $V_c$ was
the standard deviation of the uniform deviate on the constraint interval.
Corresponding to this best $V_c$ were the three best fit parameters
\begin{figure}
\epsfxsize =3in
\centerline{\epsffile{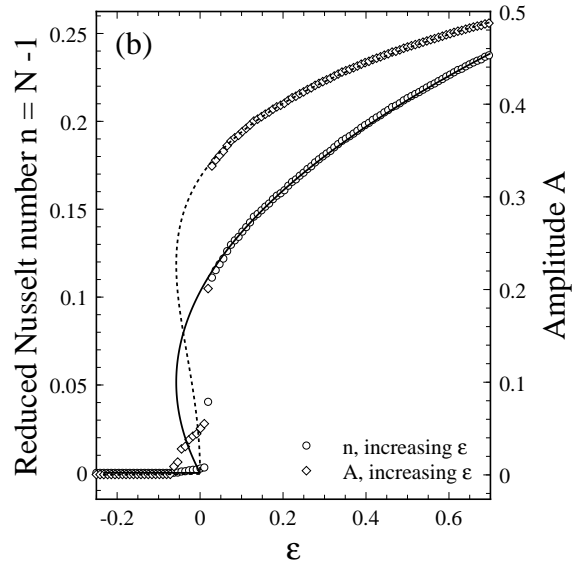}}
\vskip 0.15in
\caption{The amplitude 
$A$ and the reduced Nusselt number $n=A^2$ {\it vs.} 
the control parameter $\epsilon$ for radius
ratio $\alpha = 0.80$ with an applied shear.  The solid and
dashed lines are a fit to the Landau amplitude equation.}
\label{subcritical_plot}
\end{figure}
\vskip 0.15in

\noindent $g$, $h$ and $f$.  The uncertainties in $g$, $h$ and $f$ were estimated by a Monte Carlo
data decimation step with $V_c$ constrained to its best fit value.

For several reasons, it was necessary to restrict the fit to the neighborhood of $\epsilon \sim 0$.  
In the first place, amplitude equation models are only rigorously valid in the limit $\epsilon \ll
1$, although in practice they have been found to apply over a more extended range.\cite{smao_stuff}
Secondly, restricting the range of $\epsilon$ reduced the impact of the residual, uncorrected
component of the conductivity drift.  In many cases, the drift effects were sufficiently large that only the data acquired with increasing $\epsilon$ were fit.

Figure~\ref{supercritical_plot} shows a typical result of transforming current-voltage data to reduced Nusselt numbers $n$ and amplitudes $A$ and fitting to Eqn.~\ref{amplitudemodel}. The transition from conduction to convection is
continuous and the best fit parameter $g > 0$,  indicating that the bifurcation is supercritical.  Figure~\ref{subcritical_plot} shows the analogous result for a
film under a strong shear. For this case, the transition from conduction to convection is discontinuous and $g < 0$; the bifurcation is subcritical.  In all fits, we found $h > 0$, and $0 < f \ll 1$.  We performed a large number of such fits and surveyed the dependence of $g$ on $\alpha$, ${\cal P}$, and
${\cal R}e$. In addition, the fits provided determinations of the critical voltage $V_c$ as a function of $\alpha$, ${\cal P}$, and ${\cal R}e$ {\it via} the bootstrap process described above.  These results are discussed in Sections~\ref{gRe=0} and \ref{gRe} below.

\subsection{Tests of linear stability theory}
\label{comparisonlinear}

This section compares the experimental measurements with the predictions of the linear stability theory given in Ref.~\cite{DDM_pof_99}.

The simplest feature of linear theory that can be compared with experiment concerns the critical
mode number for zero shear, $m_c^0$.  As mentioned in Section~\ref{experiment}, some experiments
were performed in films with slight thickness  nonuniformity. This permitted the qualitative
visualization of the flow field and a quantitative determination of the mode number $m$ of the
stationary pattern of vortices. Unless the bifurcation is strongly subcritical, we expect $m =
m_c^0$ close to onset. The observed zero shear mode number was in excellent agreement with
predictions of linear stability analysis. Table~\ref{marginalmode} summarizes the results. Rapid
rotation of the vortex pattern and larger hysteresis prevented a systematic study of $m$ under
shear.  Qualitative observations confirm the general prediction that $m_c({\cal R}e > 0) <
m_c^0$, {\it i.e.} that shear reduces the number of vortices.

\begin{figure}
\epsfxsize =3.2in
\centerline{\epsffile{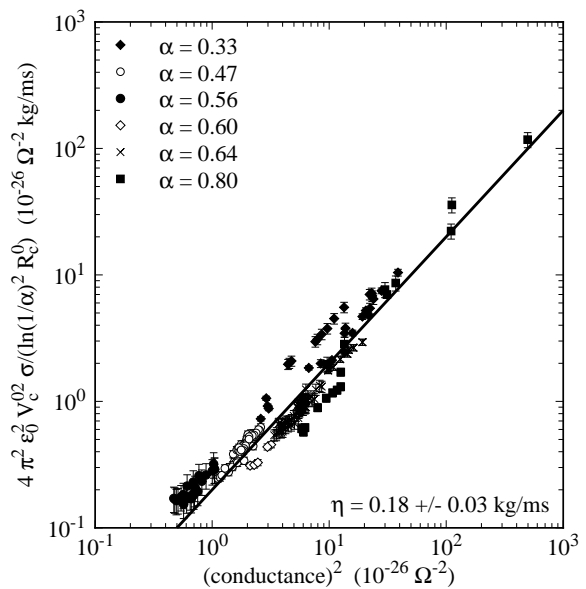}}
\vskip 0.15in
\caption{The scaled critical voltage {\it vs.} the square of the conductance for 
various films at
six different $\alpha$.  The solid line is a one-parameter fit with a single
 constant of proportionality 
which determines the viscosity $\eta = 0.18 \pm 0.03$
kg/ms.}
\label{scaledVc2_vs_cond2}
\end{figure}

The primary theoretical result of Ref.~\cite{DDM_pof_99} is the prediction of the
critical voltage $V_c$ required for the onset of electroconvection. $V_c$ is given for the general case by Eqn.~\ref{RcVc}. Even though the viscosity $\eta$ is not well-known, we can test the zero shear theory for various thicknesses $s$ and conductivities $\sigma$ using the following scheme. Denoting the zero shear value of $V_c$ by $V_c^0$, we write
\begin{equation}
({V_c^0})^2(\alpha) =\Biggl[\frac{ \sigma \eta
s^2}{{\epsilon_0}^2}\Biggr]{\cal R}_{c}^0(\alpha) \,. \label{VcRe0}
\end{equation}
Using Eqn.~\ref{conductance}, Eqn.~\ref{VcRe0} can be
expressed more conveniently as
\begin{equation}
v^2 = \frac{{4 \pi^2 \epsilon_0}^2 \sigma ({V_c^0})^2(\alpha)}{{(\ln( 1/\alpha
))}^2{\cal R}_{c}^0(\alpha)} = \eta c^2\,. \label{scaledVc}
\end{equation}
Written in this way, Eqn.~\ref{scaledVc} expresses a proportionality between $v^2$ and $c^2$ in which the viscosity $\eta$ is the only unknown parameter.  Consistency with this proportionality over a wide range of parameters serves as a test of the linear theory for $V_c^0$ as well as a determination of $\eta$.  There is one caveat: this analysis is not entirely experimental but requires the theoretical value of ${\cal R}_{c}^0(\alpha)$.  The quantity $v$, which we refer to as the scaled critical voltage, was found as follows.  For each set of  current voltage data,  the fit procedure outlined in Section~\ref{dataanalysis} was used to deduce a critical voltage $V_c^0$ and conductance $c$ at onset.  The film thickness $s$ was deduced from the color of the film.  Using the measured radius ratio $\alpha$ and Eqn.~\ref{conductance},  $\sigma$ was
calculated. Finally, the numerical result for ${\cal R}_{c}^0(\alpha)$ was used to find $v^2$ which was plotted against $c^2$. Figure~\ref{scaledVc2_vs_cond2} shows the results obtained from $228$ current-voltage runs at
six different $\alpha$, numerous different $s$ and  
conductivities in the range $ 5.9 \times 10^{-8} < \sigma < 8.4 \times 10^{-7}
{\Omega}^{-1} \rm {m}^{-1}$. Consequently, the range of
${\cal P}$ is very broad.  
\vskip 0.2in

\noindent Within some scatter,  Fig.~\ref{scaledVc2_vs_cond2} exhibits the
predicted  
proportionality over several decades. Hence, the theory properly accounts for the scaling of the critical voltage with respect to the film thickness and radius ratio.

A single-parameter linear fit to
$v^2 = \eta c^2$  gives $\eta = 0.18 \pm 0.03$ kg/ms.  This is a reasonable value for the viscosity; while it has not been independently measured, it is
believed to be of order of $0.1$ kg/ms.\cite{viscosity2}

We now turn to the case of nonzero shear.
The main prediction of the linear stability analysis is that the onset of electroconvection is suppressed by Couette shear.  This suppression is both $\alpha$ and ${\cal P}$ dependent. Returning to Eqn.~\ref{eptilde}, the degree of suppression $\tilde{\epsilon}$ is given by
\begin{equation}
\tilde{\epsilon}~(\alpha,{{\cal R}e},{\cal P}) = \Biggl[\frac{{\cal
R}_c(\alpha,{{\cal R}e},{\cal P})}{{\cal R}_c^0(\alpha)}\Biggr] -1 =
\Biggl(\frac{V_c(\alpha,{\cal R}e)}{V_c^0(\alpha)}\Biggr)^2 - 1\,.
\label{eptilde2}
\end{equation}
We used the two equivalent expressions for $\tilde{\epsilon}$ in
Eqn.~\ref{eptilde2} to calculate the suppression theoretically and
experimentally.  $V_c(\alpha,{\cal R}e)$ was found from the nonlinear fit, along with the
conductance $c$ for that particular sweep.  It was necessary to correct $V_c^0(\alpha)$ for the
drift of $c$ in order to calculate $\tilde{\epsilon}$. This was done using values of $V_c^0(\alpha)$ taken from zero shear sweeps 
performed before and after each sheared sweep.  The variation of $V_c^0(\alpha)$ with $c$ was modelled with a linear fit which was 
used to find the drift-corrected value of $V_c^0(\alpha)$. In this way each experimental value of $\tilde\epsilon$ could be be 
computed for a constant $c$.  The uncertainty in  $\tilde\epsilon$ was dominated by the uncertainty 
\begin{figure}
\epsfxsize =3in
\centerline{\epsffile{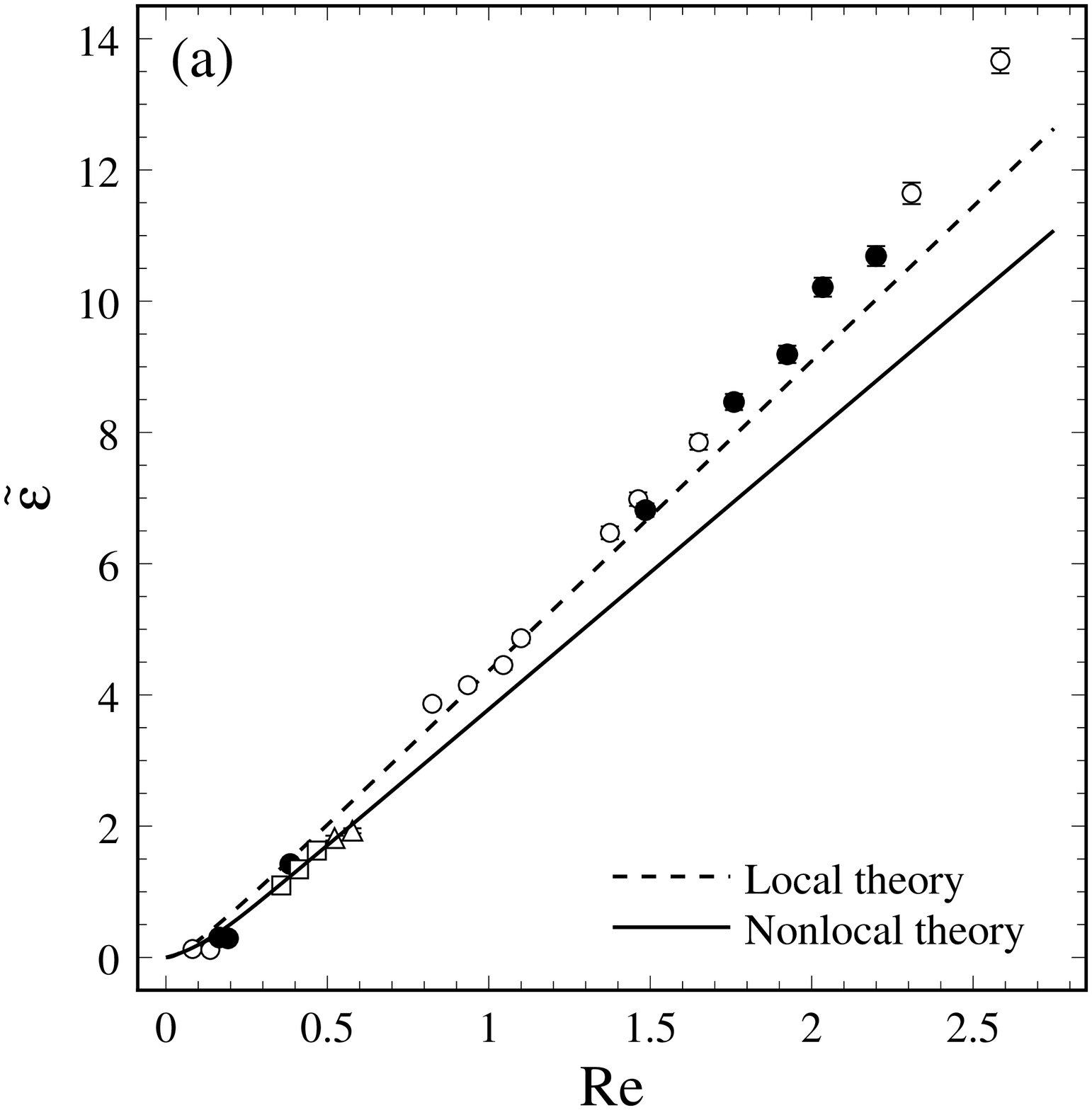}}

\epsfxsize =3in
\centerline{\epsffile{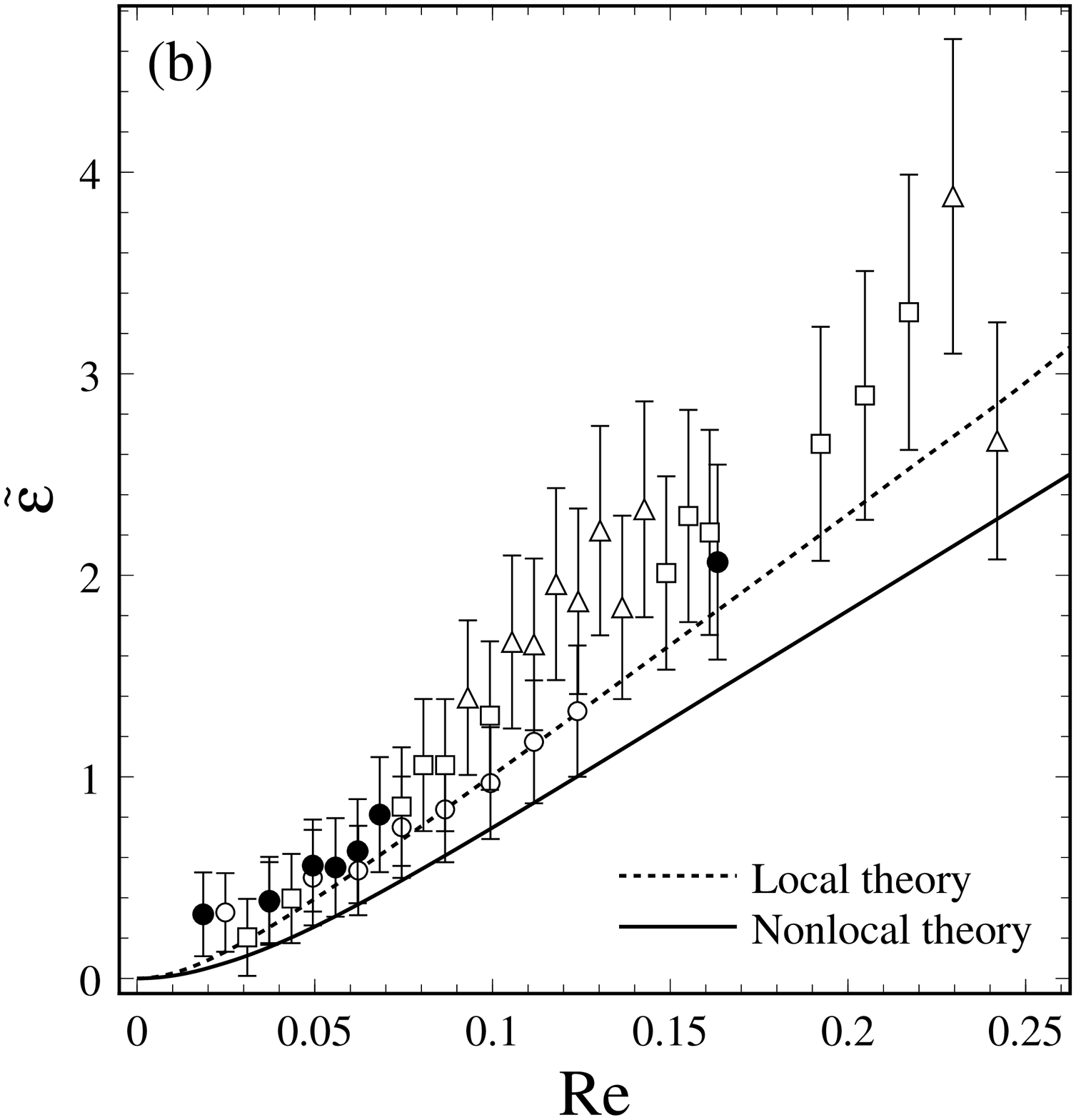}}
\caption{A comparison between the measured
suppression $\tilde\epsilon$ {\it vs.} ${\cal R}e$ and the predictions of local
and nonlocal theory based on Ref.
[3].
 In (a) $\alpha
= 0.47$.  The different symbols denote the ${\cal P}$-quartiles: 
$13.3 < \circ  < 15.4 < \bullet < 17.5 < \Box < 19.6 < \triangle < 21.7$.
The theoretical lines are for the mean ${\cal P} =16.3$ of the data. Similarly, 
in (b) $\alpha = 0.64$, $29.1 < \circ  < 37.1 < \bullet < 45.2 < \Box < 53.2 < 
\triangle < 61.2$ and the theoretical lines are for the mean ${\cal P} = 45.2$. Note the very different scales.}
\label{suppression_plot}
\end{figure}
\vskip 0.15in

\noindent in the drift correction 
to $V_c^0$.  The conductance $c$ for each $\tilde\epsilon$ can be used to calculate its associated value of ${\cal P}$.  
Thus, we arrive at sets of experimental values of $\tilde\epsilon(\alpha,{{\cal R}e},{\cal P})$ for fixed $\alpha$ and ${{\cal R}e}$ and 
ranges of ${\cal P}$.  These can be compared in detail to the theoretical results for suppression as a function of these parameters 
given in Ref.~\cite{DDM_pof_99,zahir_phd_thesis}.  Figures~\ref{suppression_plot}a and b show this comparison at $\alpha = 0.47$ and $\alpha = 0.64$ 
respectively.  In each case the  theoretical curves are calculated for the mean value of the range of ${\cal P}$ spanned by the data.  
For a discussion of the distinction between the local and nonlocal approximations, see Ref.~\cite{DDM_pof_99}. 

Note that the ranges of ${\cal R}e$ for these two $\alpha$ are different by a factor of ten and that the suppressions are also very different.  It is quite interesting that a rather mild shear (${{\cal R}e} \sim 3)$ suppresses the onset of electroconvection so strongly that $\tilde\epsilon \sim 14$ ({\em i.e.} ${\cal R}_c = 15{\cal R}_c^0$). We have also studied the suppression at $\alpha = 0.33, 0.56, 0.60$, and $0.80$. The results are similar to Fig.~\ref{suppression_plot}. The suppression  data covers the range $10 < {\cal P} < 130$ and $0 \leq {\cal R}e < 3$.  In each case the agreement between theory and experiment is fair, with the theory slightly underestimating the suppression in most cases.  The upper bound on ${\cal R}e$ could easily be extended using increased rotation rates, while smaller ${\cal P}$ would require larger, thicker films.

The degree of agreement shown in Figs.~\ref{suppression_plot} is essentially independent of the value of $\eta$.  
Recall that $\eta$ was determined by a single parameter fit to Eqn.~\ref{scaledVc}.  Since the $\eta$ dependence in
the ${\cal R}e$ scaling of both the theory and the experiment are proportional to $1/\eta$, any change in $\eta$
multiplies both by the same factor.\cite{DDM_pof_99} This simply results in a rescaling of the ${\cal R}e$ axis in
Figs.~\ref{suppression_plot}a and b, with no change in the quality of the agreement.

We close this section with some brief speculation on what might account for the scatter in
Fig.~\ref{scaledVc2_vs_cond2} and for the remaining discrepancies between theory and experiment in the suppression
curves.

The most likely sources of these discrepancies are the imperfect geometry of the film and its
finite thickness.  The 2D theoretical model assumed that the flow velocity is independent of
position over the thickness of the film.  This may be inexact for thicker films.  Since the
electrical forcing is localized near the free surfaces, it seems likely that the surfaces are
preferentially driven in thick films so that the motion is not accurately 2D.  The film's edges
are also imperfect because small wetting layers unavoidably exist on the circumferences of the
inner and outer electrodes.  These may produce electrical or velocity boundary conditions that
are not exactly those assumed by the theory.  In general, however, the linear stability theory
works remarkably well considering its rather simple assumptions.

\subsection{Coefficients of the Cubic Nonlinearity without Shear}
\label{gRe=0}

In this section, we consider the experimental results in the nonlinear regime, beginning with
the unsheared ${\cal R}e = 0 $ case. These are expressed in terms of the cubic Landau
coefficient $g$ of the amplitude equation, Eqn.~\ref{amplitudemodel}.  Even with ${\cal R}e$
fixed at zero, we will show that this coefficient is an interesting and nontrivial function of
the remaining dimensionless quantities, ${\cal P}$ and $\alpha$.

The radius ratio $\alpha$ is a fixed geometric parameter for each run, determined only by the 
choice 
\begin{figure}
\epsfxsize =3in
\centerline{\epsffile{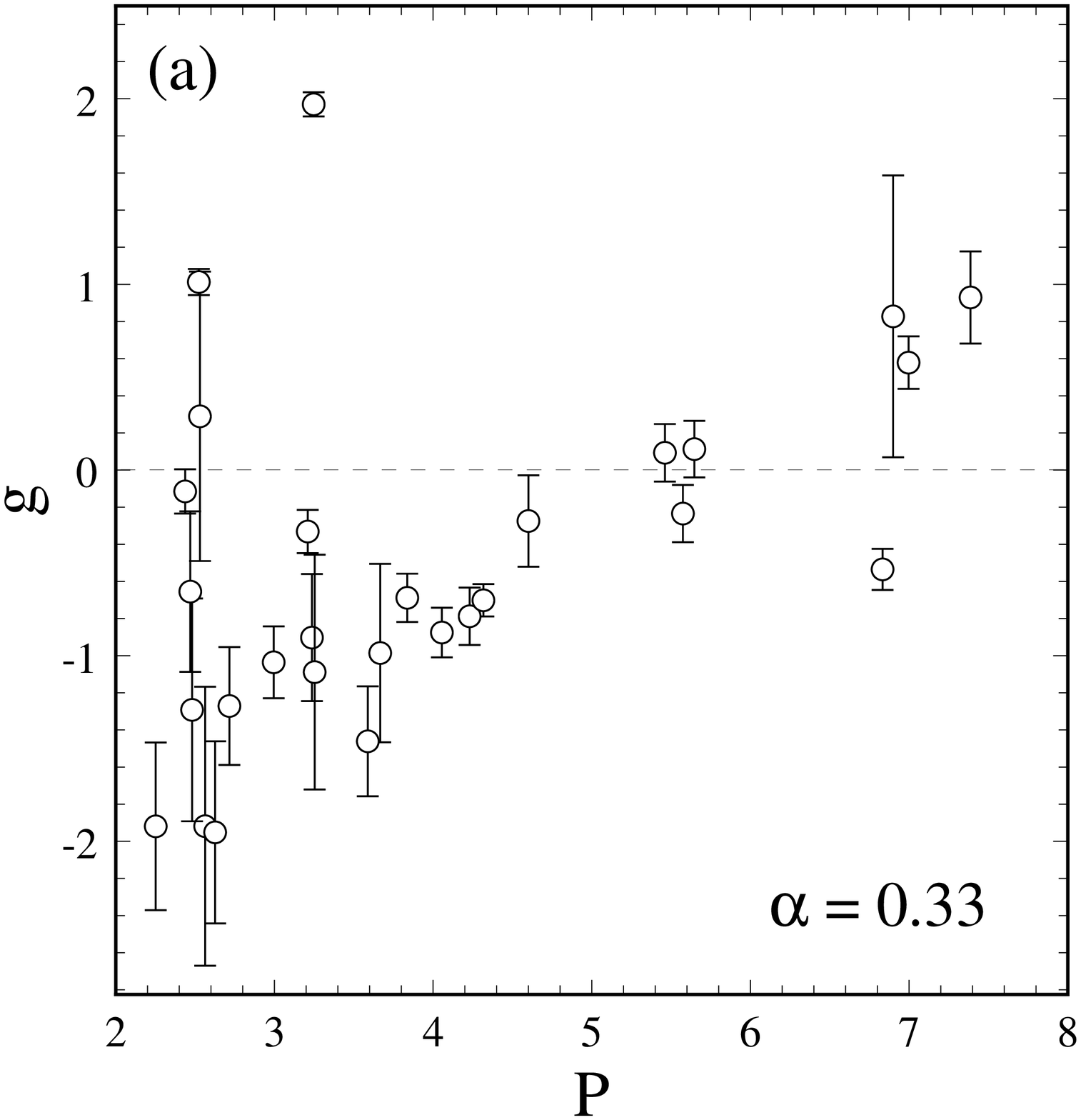}}
\epsfxsize =3in
\centerline{\epsffile{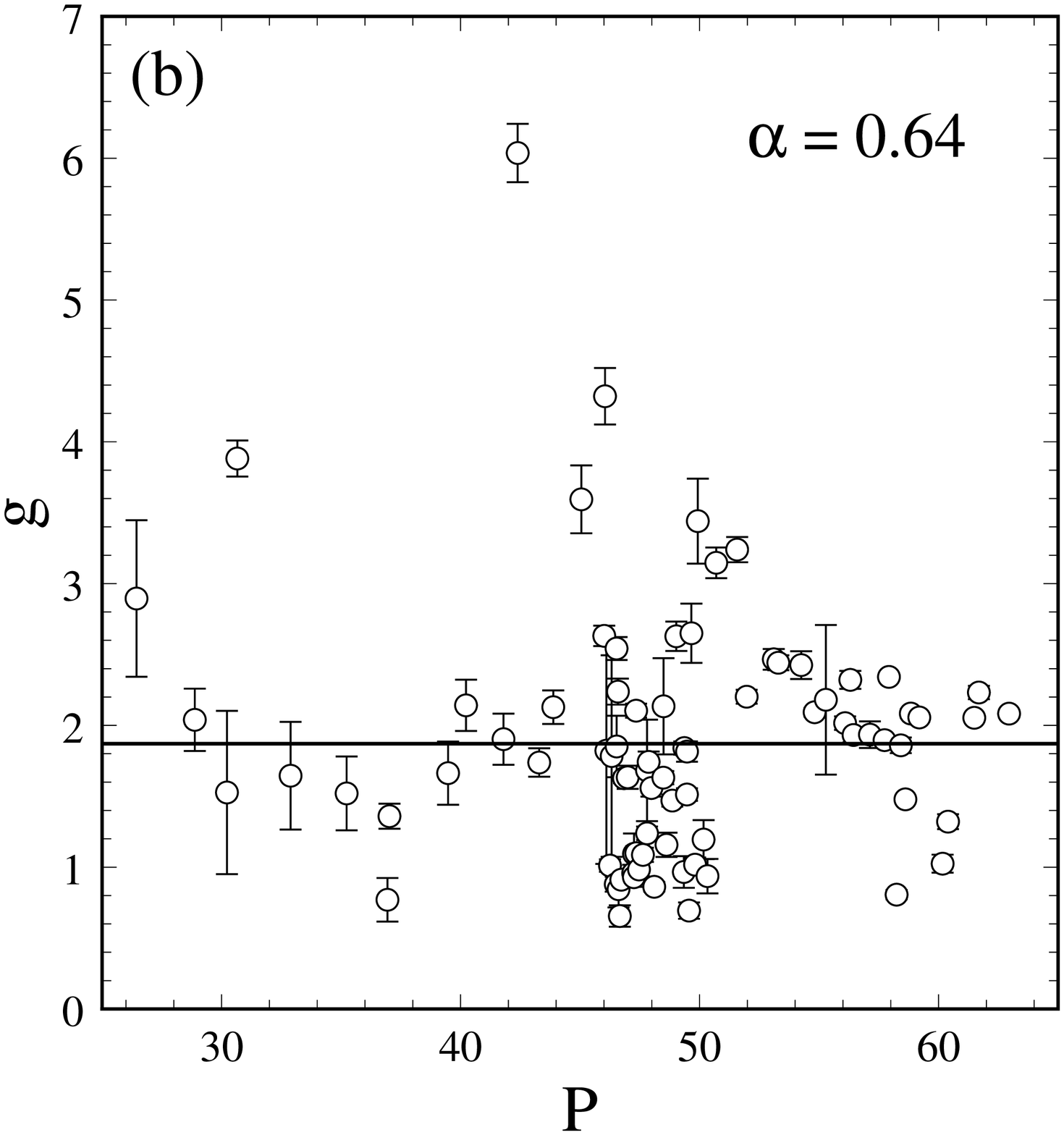}}
\vskip 0.15in
\caption{(a) Measurements of the coefficient of the cubic nonlinearity $g$ for zero shear {\it vs.} ${\cal P}$ 
for (a) $\alpha = 0.33$ and  (b) $\alpha
= 0.64$. }
\label{g_vs_P_Re=0}
\end{figure}
\vskip 0.15in

\noindent of the electrode dimensions.  
The ${\cal P}$ appropriate to each film was deduced after each run 
as described
in Section~\ref{dataanalysis}.  Since ${\cal P}$ is proportional to the conductivity, which exhibits a slow drift, a 
wide
range of ${\cal P}$ could be investigated. While they are independent in principle, $\alpha$ and ${\cal P}$ are not
easy to separate experimentally. This practical constraint derives from Eqn.~\ref{pdimensional} in
which ${\cal P} \propto d^{-1}$, where the width of the film $d = r_o - r_i$. In practice, $d$ is largest at small
$\alpha$, so the runs with the smallest ${\cal P}$ occur for small $\alpha$.

The nonlinear coefficient $g(\alpha, {\cal P})$ was determined as described in Section~\ref{dataanalysis}.
Figures~\ref{g_vs_P_Re=0}a and b show $g$ as a function of ${\cal P}$ for two different $\alpha$. 
The scatter that is manifest in these
plots exceeds the statistical uncertainty of the fit.  As discussed in
Section~\ref{dataanalysis}, the scatter originates from systematic effects
due to the non-ideal features of the experiment.  Nevertheless, the gross features of the ${\cal P}$ 
dependence of $g$ can still be extracted.

At $\alpha = 0.33$, the measurements explored the range $2 < {\cal P} < 8$. These were the smallest ${\cal P}$
reached in the experiment. It is clear from Fig.~\ref{g_vs_P_Re=0}a that $g$ increases with 
${\cal P}$, and passes through zero for small enough  ${\cal P}$. Thus, for $\alpha =
0.33$, we find that the bifurcation from the conduction state to the $m = 4$
 vortex state is subcritical ($g < 0$) for
${\cal P} \stackrel{<}{\sim} 5$ and supercritical ($g > 0$) for ${\cal P}
\stackrel{>}{\sim} 5$. Near ${\cal P}
\stackrel{\sim}{=} 5$ we pass a tricritical point. It is interesting to observe that in each case the bifurcation
involves the {\it same} two symmetry states, yet its subcriticality depends on ${\cal P}$.

For all the other, larger $\alpha$ investigated, $g$ was found to be independent of ${\cal P}$. This is illustrated
in Fig.~\ref{g_vs_P_Re=0}b for $\alpha = 0.64$ and was also true for $\alpha = 0.47,~0.56,~0.60,$~and~$0.80$.
However, for all of these cases, ${\cal P}$ was never less than $10$ and can be as large as several hundred.  Thus,
it is unclear whether the subcriticality at $\alpha = 0.33$ is due to the smallness of ${\cal P}$ or the smallness
of $\alpha$, or some combination. It would be interesting to determine whether $g$ also becomes negative for large
$\alpha$ as ${\cal P}$ decreases. The regime of large $\alpha$ and small ${\cal P}$, while accessible in principle,
would require significantly larger electrodes or much thicker films.  All electroconvection experiments on freely
suspended films in the rectangular geometry have reported supercritical bifurcations\cite{smorris1,smorris2,smorris3,smao_stuff}.  
However, these experiments were also at large ${\cal P}$ and so the possibility of a subcritical bifurcation in
rectangular films cannot be excluded.

The case of large ${\cal P}$ is easier to understand. In the governing electrohydrodynamic
equations (see Ref.~\cite{DDM_pof_99}), the parameter analogous to the Prandtl number only
appears as the inverse, ${\cal P}^{-1}$, multiplying certain nonlinear terms.  Hence, it is
perhaps not surprizing that $g$ becomes independent of ${\cal P}$ for ${\cal P}
\stackrel{>}{\sim} 10$, where these terms become negligible.

In order to examine how $g$ depends on $\alpha$, we removed the ${\cal P}$ dependence by averaging
 over data at various ${\cal P}$.  For the five largest $\alpha$, $g$ is independent of ${\cal P}$ over broad ranges
of ${\cal P}$.  A weighted average of $g$ was obtained for these cases.  For $\alpha = 0.33$, where some ${\cal P}$
dependence was evident, only $g$ values in the narrow range $2.1 < {\cal P} < 4.4$ were averaged.

\begin{figure}
\epsfxsize =3.1in
\centerline{\epsffile{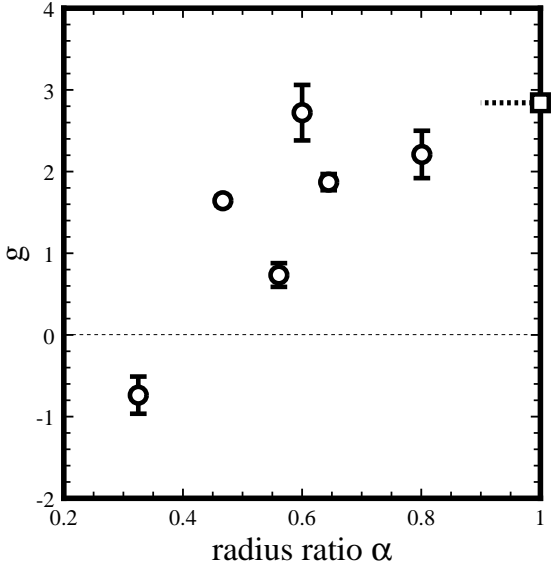}}
\vskip 0.1in
\caption{Measurements of the coefficient of the cubic nonlinearity $g$ for zero shear {\it vs.} the radius ratio $\alpha$.
The dashed line shows the theoretical value for $g$ in the limits ${\cal P} \rightarrow \infty$ and $\alpha \rightarrow 1$
from Ref.~[25].}
\label{g_vs_rr}
\end{figure}

In this averaging the systematic scatter in $g$ was treated as random.  It is thus likely that the true uncertainty
in the average values of $g$ will be much larger than the standard deviation of the mean.  The ultimate test of this
procedure lies in the comparison of the averaged values of $g$ with theoretical predictions.  As we describe below,
the limited comparison we are able to make at present is not unfavourable. The line in Fig.~\ref{g_vs_P_Re=0}b
shows the average value of $g$ for the plotted data.  All of the results for $g(\alpha)$ are tabulated in
Table~\ref{gvsrrtable} and plotted in Fig.~\ref{g_vs_rr}.

\vskip 0.1in

It is clear from Fig.~\ref{g_vs_rr} that, overall, $g(\alpha)$ increases with $\alpha$.  As discussed above, the
larger $\alpha$ data are also associated with larger ${\cal P}$.  At present, no theoretical predicion is available
for $g$ at arbitrary $\alpha$ and ${\cal P}$. However, we expect $g$ to approach a
limiting value quite rapidly as $\alpha \rightarrow 1$. This limit corresponds to an unbounded lateral geometry
in which the film is a strip of fluid suspended at its long parallel edges by two semi-infinite plate electrodes.
In this limit the discrete azimuthal mode numbers $m$ are replaced by a continuous wavenumber $k=m/{\bar r}$, 
where ${\bar r} = (r_i + r_o)/2$ is the mean radius.
From linear theory, we found that the critical parameters ${\cal R}_c^0$ and 
$k_c^0 = m_c^0/{\bar r}$ at $\alpha = 0.80$ are
already very close to the limiting values for $\alpha \rightarrow 1$.\cite{DDM_pof_99}

This limiting behaviour is reasonable given the large dimensionless circumference for large $\alpha$. The relevant
aspect ratio $\Lambda$ is the circumference at the mean film radius, $\pi (r_i + r_o)$, divided by the width of the
film $d = r_o - r_i$, so that $\Lambda = \pi(1 + \alpha)/(1 - \alpha)$.  At $\alpha = 0.80$, $\Lambda \sim 28$.  
Almost all the experiments performed in the rectangular geometry had $\Lambda \stackrel{<}{\sim}
10$~\cite{smorris1,smorris2,smorris3,smao_stuff,endselection} and were well-modelled by the theory for an unbounded strip for
which $\Lambda = \infty$.  It is thus resonable to expect $g$ at $\alpha = 0.80$ to be close to its limiting value
for $\alpha = 1$.  Similarly, as the large $\alpha$ data also involve very large values of ${\cal P}$, we may also
employ the theory in the ${\cal P} \rightarrow \infty$ limit.

Weakly nonlinear analysis of unsheared electroconvection in the `plate' electrode geometry was presented in
Ref.~\cite{gle97} for the case ${\cal P} = \infty$. The result of that analysis, $g = 2.842$ is shown
in Fig.~\ref{g_vs_P_Re=0}. It is in good agreement with a reasonable extrapolation of the data to the
$\alpha \rightarrow 1$ limit.

\begin{figure}
\epsfxsize =3.2in
\centerline{\epsffile{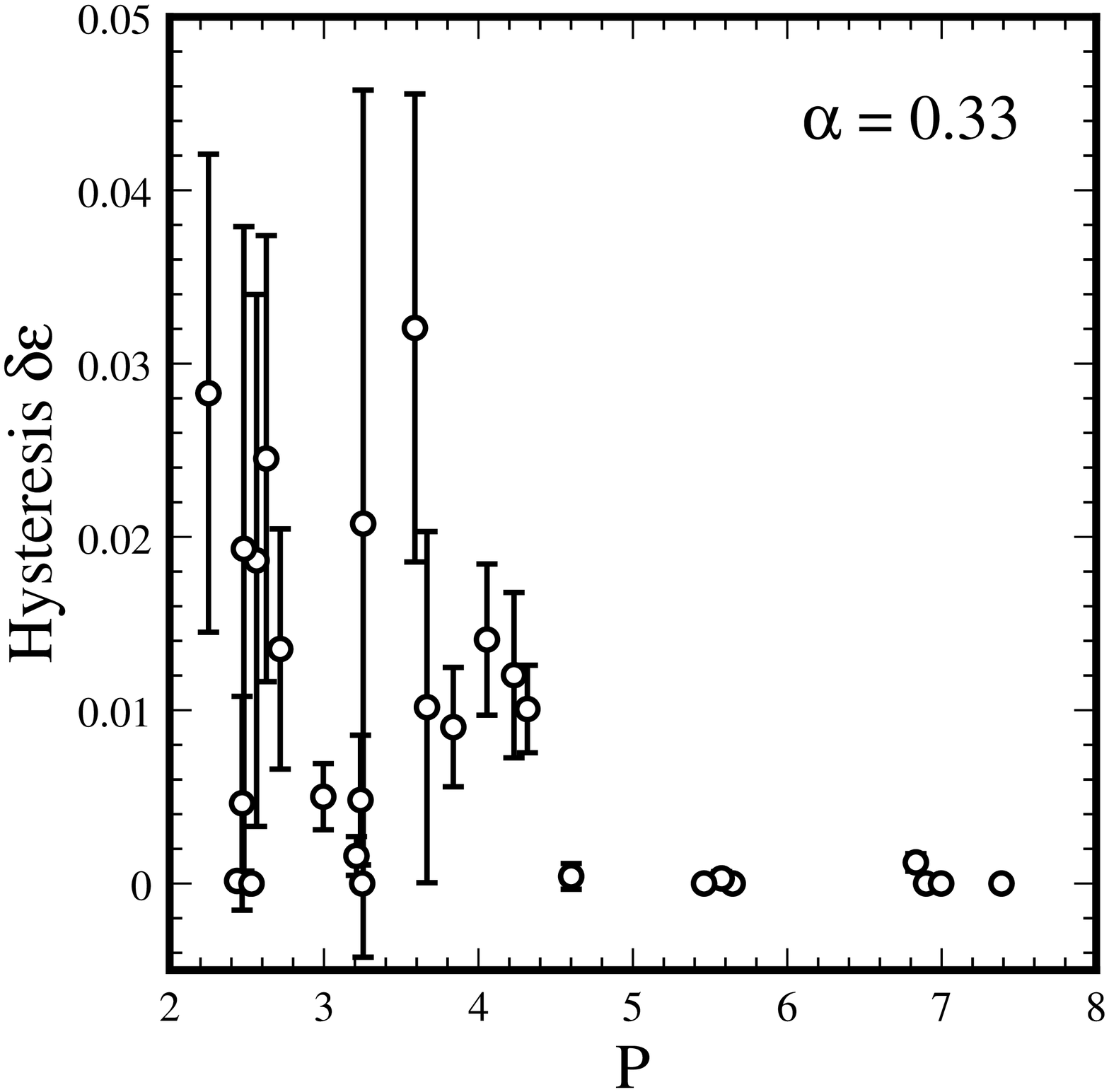}}
\vskip 0.15in
\caption{Measurements of the hysteresis width $\delta \epsilon$ {\it vs.} ${\cal P}$} for radius ratio $\alpha = 0.33$.
\label{hyst_vs_P}
\end{figure}
\vskip 0.15in

When $g < 0$, the onset of convection becomes hysteretic and the quintic term with 
coefficient $h$ in Eqn.~\ref{amplitudemodel} becomes
significant.\cite{zahir_phd_thesis} Since $0 < f \ll 1$, the width of the hysteresis loop $\delta\epsilon$ in
$\epsilon$ is given by $\delta\epsilon = g^2/4h$.  When $g \geq 0$, $\delta\epsilon = 0$. The
hysteresis width with zero applied shear is plotted in Fig.~\ref{hyst_vs_P} for $\alpha = 0.33$ and various ${\cal
P}$. Note that the hysteresis vanishes for ${\cal P} \stackrel{>}{\sim} 5$ and even when non-zero, is always
rather small, $\delta\epsilon \sim 0.02$. This is in contrast with the sheared case, discussed in the next Section, for
which $\delta\epsilon \sim 0.1$.

\subsection{Coefficients of the cubic nonlinearity with shear}
\label{gRe}

In the presence of shear, the coefficient $g$ of the cubic nonlinearity is strongly dependent
on the Reynolds number ${\cal R}e$ of the Couette flow imposed in the base state.  We find
that the shear drives $g$ negative, producing a hysteretic Hopf bifurcation to convection in
the form of travelling vortices.  In this Section, we describe the ${\cal R}e$ dependence of
$g$, which, as in the previous Section, is also a function of the radius ratio $\alpha$ and
the dimensionless ratio ${\cal P}$. 

\begin{figure}
\epsfxsize =3.2in
\centerline{\epsffile{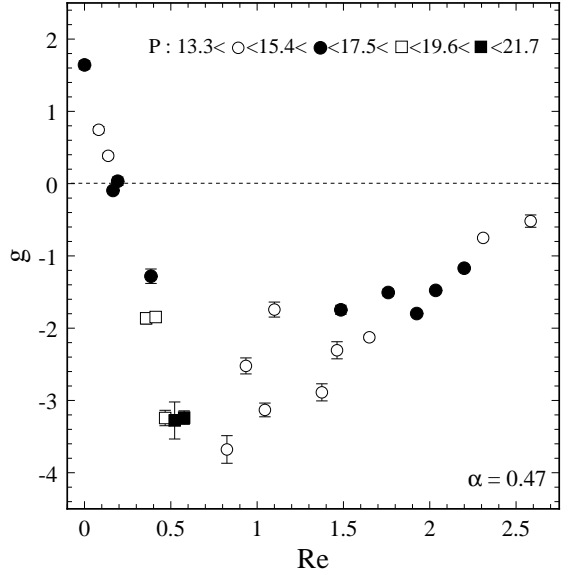}}
\vskip 0.15in
\caption{Measurements of the coefficient of the cubic nonlinearity $g$ {\it vs.} the Reynolds number 
${\cal{R}}e$ at $\alpha = 0.47$. The different symbols denote
data in various quartiles of ${\cal P}$.}
\label{g_vs_Re_rr=0.47}
\end{figure}

Figure~\ref{g_vs_Re_rr=0.47} shows a representative example of the ${\cal R}e$
dependence of $g$, in this case for $\alpha = 0.47$. At ${\cal R}e = 0$, the value of $g$ is found, as described in
Section~\ref{gRe=0}, by averaging over a range of ${\cal P}$, which was always $> 10$.  In the case
shown in Fig.~\ref{g_vs_Re_rr=0.47}, $g({\cal R}e=0) = 1.64 \pm 0.06$ with a mean ${\cal P}$ of
$16.3$.  The plotted $g$ data for ${\cal R}e > 0$ had $13.3 < {\cal P} < 21.7$.

Our main result is that ${\cal R}e$ plays the part of a second control parameter 
which allows the primary
bifurcation 
\vskip 0.2in

\noindent to electroconvection to be tuned between super- and subcritical.  For $\alpha = 0.47$, the
bifurcation is supercritical at ${\cal R}e = 0$ and weakens as the Reynolds
number increases.  The bifurcation becomes tricritical at ${\cal R}e \sim 0.2$ and
is subcritical thereafter.  The subcriticality at first deepens with increasing Reynolds number, until 
at ${\cal R}e \sim 0.85$ a minimum value of $g
\stackrel{\sim}{=} -3.7$ is reached.  For ${\cal R}e > 0.85$, the bifurcation remains
subcritical but $g$ becomes an increasing function of the Reynolds
number.  For the range of Reynolds numbers investigated the bifurcation does
not become supercritical again.

There is some scatter in the data, nonetheless, the overall trends are
clear. The systematic deviations are
comparable to those in Figs.~\ref{g_vs_P_Re=0}a and b.  The results were  
qualitatively similar for $\alpha = 0.56, 0.60, 0.64,$ and $0.80$. The tricritical Reynolds number 
at which $g = 0$ will be denoted ${{\cal R}e_T}$.   The value of ${{\cal R}e_T}$  varys for different $\alpha$ 
and ${\cal P}$, 
but the general features of the $g$ {\it vs.} ${\cal R}e$ curve are preserved over a wide range of parameters. 
Table~\ref{Reg=0} lists the values of
${\cal R}e_T$ under various conditions.

The variation of ${\cal R}e_T$ is probably due to the combination of changing 
 both $\alpha$ and ${\cal P}$, rather than to the variation either parameter separately. 
These are difficult to  experimentally disentangle and the parameter space is large.
For zero shear, $g$ was
found to be independent of ${\cal P}$ for ${\cal P} \stackrel{>}{\sim} 10$, but under shear
there is insufficient data to draw many conclusions about the ${\cal
P}$ dependence of $g$.  If linear theory is any
guide, we expect there will in general be a
greater ${\cal P}$ dependence in the sheared case.  It was established in Ref.~\cite{DDM_pof_99}
that the linear
theory result for ${\cal R}_c$ was independent of ${\cal P}$ for ${\cal R}e = 0$, 
\begin{figure}
\epsfxsize =3.2in
\centerline{\epsffile{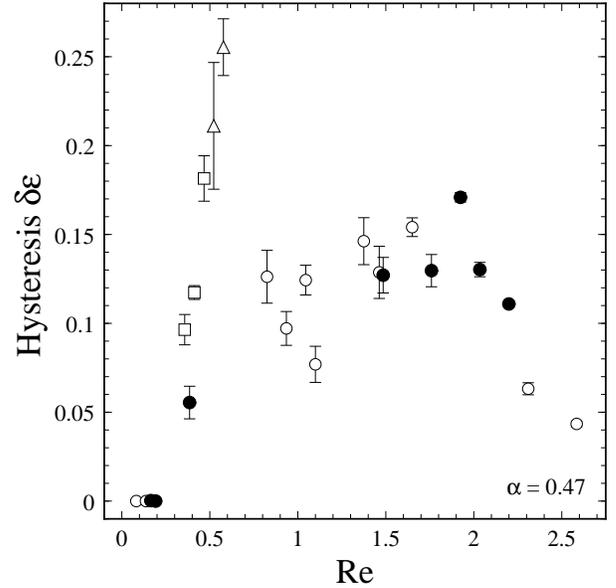}}
\vskip 0.15in
\caption{The hysteresis width $\delta \epsilon$ {\it vs.} the Reynolds number ${\cal R}e$ for radius ratio $\alpha = 0.47$. The symbols denote the same quartiles of ${\cal P}$ as in Fig.~11.}
\label{hyst_vs_Re_rr=0.47}
\end{figure}
\vskip 0.2in

\noindent but weakly dependent 
on ${\cal P}$ for ${\cal R}e \neq 0$.
 It appears from the tabulated results that as
${\cal P}$ and $\alpha$ decrease, ${\cal R}e_T$ increases.

The minimum value of $g$ as a function of ${\cal R}e$ also vary for different $\alpha$ and ${\cal P}$. 
We will refer to the coordinates of this special point as $g_{min}$ and ${\cal R}e_{min}$.   
Table~\ref{ming} lists the minimum values observed.  A look at Fig.~\ref{g_vs_Re_rr=0.47} shows that the minimum
value is only known up to the density of data and to within the scatter in the whole plot.  
The minimum is a local one and is observed within ranges of ${\cal R}e$ which had different maxima for
different $\alpha$.  

Since the parameter space for the data is defined by several parameters it
becomes difficult to make meaningful comparisons of the experimental results
when more than one parameter changes.  However, one fortunate comparison can be
gleaned from Table~\ref{ming}.  At $\alpha = 0.47$, the film had ${\cal P} =
15.3$ while at $\alpha = 0.80$, ${\cal P} = 12.0$.  Since the radius ratios
are very different and the ${\cal P}$ are not, it is not unreasonable to
directly compare the values of $g_{min}$ and ${\cal R}e_{min}$ for these
cases.  It is evident that the bifurcation is much more strongly subcritical
at $\alpha = 0.80$ than at $\alpha = 0.47$, and the values of ${\cal R}e_{min}$ are also very different.

The coefficient of the quintic nonlinearity $h$ as function of ${\cal R}e$ is studied in
Ref.~\cite{zahir_phd_thesis}. Figure~\ref{hyst_vs_Re_rr=0.47} shows the the hysteresis width $\delta\epsilon =
g^2/4h$ as a function of ${\cal R}e$ for $\alpha=0.47$. The hysteresis when ${\cal R}e > 0$ is much larger than that
described in the previous Section which was found for ${\cal R}e = 0$ at small $\alpha$ and ${\cal P}$.

The Nusselt number of the convection is independent of the Hopf frequency ({\it i.e.} the
travelling rate of the vortex pattern). It would be interesting to measure this frequency,
which is governed by the imaginary part of the complex Ginzberg-Landau equation,
Eqn.~\ref{AMP6_imag}, and depends on $\epsilon$, $\alpha$, ${\cal R}e$ and ${\cal P}$.  This
would however require visualization of the flow pattern or some spatially localized probe of
the amplitude.

\section{Discussion}
\label{discussion}

Because of its rather simple symmetry and forcing scheme, annular electroconvection under 
applied shear has many similarities to other systems and models.  In this section, we discuss 
these similar systems with a view to putting annular electroconvection into a general context. 

Electroconvection is obviously analogous to buoyancy-driven thermal convection.  In fact, it can be shown that
the linear stability problem for surface charge driven electroconvection in thin films reduces to the 
2D Rayleigh-B\'{e}nard problem if the nonlocal coupling between fields and charges is neglected.\cite{linear}  
Detailed calculations show that this ``local'' approximation is an accurate one, even when applied in the 
weakly nonlinear regime. 

There have been some theoretical studies of 3D Rayleigh-B\'{e}nard convection (RBC)
in the presence of a {\it plane} Couette shear flow\cite{fk88}.  The theory assumed the
usual theoretical geometry of a fluid layer confined between infinite, perfectly
conducting horizontal planes. Our annular system approaches the plane one in the $\alpha \rightarrow 1$ limit.
 (The $\alpha < 1$ annular geometry  has finite
 extent and curvature; the implications of these are discussed below.)
Linear stability analysis of a plane Couette base state to RBC reveals
stability differences between transverse roll disturbances (with axes
perpendicular to the shear flow), and longitudinal roll disturbances (with
axes parallel to the shear flow). Longitudinal-roll disturbances have
identical stability properties to unsheared RBC, and are always
more unstable than the transverse-roll disturbances.  In fact, 
longitudinal-roll disturbances have stability properties that are independent
of {\it any} uni-directional shear flow along the axis of the rolls. 
In our 2D system, these more unstable longitudinal roll modes do not exist and the vortices we see 
correspond to truly 2D transverse rolls. Thus, we observe the analog of a state that would normally be preempted
by longitudinal rolls if the geometry were not constrained.

According to linear theory, transverse-roll disturbances in unbounded, sheared RBC, 
like our vortices, exhibit suppression, or added
stability due to the shear, under plane Poiseuille or plane Couette or any mixture
of these two flows.  The onset Rayleigh number for transverse rolls is a
monotonically increasing function of the shear Reynolds number, similar to
what we found for electroconvection.   Furthermore, the critical
wavenumber of the most unstable transverse disturbance was found to be a
monotonically decreasing function of the shear Reynolds number, which is analogous to the reduction 
of $m_c$ by shear that we observed.  Transverse rolls also travel under unidirectional plane Couette shear,
again analogous to our travelling vortex state.

Whereas unbounded RBC with plane Couette shear cannot be
realized experimentally, RBC has been studied experimentally and theoretically in narrow slots 
with open through-flows.\cite{ch93,rbc_flow1,rbc_flow2} Quasi-2D transverse rolls can be stabilized by wall 
effects in slots. 
The through-flow consists of a weak Poiseuille flow with a very small Reynolds number. Its
effects on RBC are well understood.  In brief, the onset of convection is
again suppressed, but the first instability is {\it convective} ({\it i.e.}
it grows only downstream of a localized perturbation), rather than {\it
absolute}.
The resulting convection pattern drifts in the direction of the through flow.
 It is interesting that the clear distinction between convective and
absolute instability is blurred in annular electroconvection with shear, in
which the `through' flow loops back on itself.  The annular geometry is
naturally {\em closed}. The linear stability analysis of Ref.~\cite{DDM_pof_99}
only treated absolute instablity. In  principle, the system might still be 
considered to be convectively 
unstable if localized perturbations
grew only as they travelled azimuthally.

Taylor Vortex flow (TVF)\cite{ch93} is an extensively studied pattern-forming instability with 
geometric similarities to annular electroconvection. However, the instability leading to
TVF depends crucially on axial disturbances to 3D Couette flow. Purely 2D
circular Couette flow is, in fact, linearly stable.\cite{drazin,wmkg95}  TVF is the result of 
an instability of a 3D shear flow,
while what we have studied here is the effect of a shear flow on the
electroconvective instability. 

Agrait and
Castellanos  theoretically studied the effect of a 3D Couette shear on an
electrohydrodynamic instability in TVF geometry.\cite{agrait88}
Their electrohydrodynamic system consisted of a nearly insulating fluid confined
between metallic electrodes.  Charge injection, a process by which charge
carriers are created at the electrodes, occurs when strong electric fields
are applied.  The interaction of this volume charge density with the
applied electric field  leads to electroconvection instabilities.\cite{melcher,felici,castellanos}
Agrait and Castellanos considered electroconvection due to a radial field with charge injection on
either cylinder. Both cylinders were permitted to rotate to produce a general
Couette shear. They found that shearing enhanced the instability,
leading to a 3D flow that resembles TVF.  This is in direct contrast to the stabilizing
effects we observed in 2D.

Although shear and rigid rotation are conceptually quite different, it is interesting to compare their
 effects on instabilities.  Again, we find crucial distinctions between 
2D and 3D systems.

The added stability in sheared annular electroconvection is a consequence of
the shear and not of rotation.  Under rigid rotation, where the inner and
outer electrodes are co-rotating, one can transform to rotating co-ordinates
in which the electrodes are stationary.  This transformation introduces a
Coriolis term $-2\Omega \hat{\bf z} \times {\vec{\bf v}} = -2\Omega \nabla
\psi $ in the Navier-Stokes equation which may be absorbed into the pressure
gradient term
$\nabla P$ and eliminated.\cite{DDM_pof_99,alonso95} Thus, in a purely 2D system,
rigid rotation and the non-rotating, unsheared case have identical stability.
 It also follows, since the transformation is general and the unsheared
bifurcation is stationary, that the resulting nonlinear vortex pattern above
onset must be stationary in the co-rotating frame.

This lack of dependence on rigid rotation may be contrasted with a large
class of 3D and quasi-2D rotating Rayleigh-B{\'e}nard
systems\cite{knobloch,goldstein93,goldstein94,rotRBC1,rotRBC2,rotRBC3}, where
rotation produces added stability but the absence of strictly 2D flow results
in a time-dependent (precessing) convection pattern in the co-rotating
frame.  The precession seen in these systems is analogous to the travelling of the sheared patterns 
that we observed only to the extent that rotation and shear break the 
same symmetry, which allows travelling patterns.  Symmetry aside, the physical origins of
added stability and precession
are fundamentally different between the 2D sheared and 3D rotating cases.

 In general, 3D rotating convecting systems may also support strictly 2D 
`Taylor column'\cite{taylor_columns} solutions which  do not precess in the rotating frame and
whose onset occurs
at the same critical Rayleigh number as in the absence of
rotation.\cite{knobloch,alonso95} The system most analogous to ours is the interesting but
experimentally unrealizable situation of 2D RBC in a rotating annular
geometry with purely {\em radial} gravity and heating\cite{alonso95}.
Theoretical studies found columnar solutions which are very closely analous to our vortices. 
One might hope to approach this limit in experiments using
radially temperature gradients imposed between rapidly rotating concentric cylinders.
Purely columnar
solutions have not observed
in any rotating RBC experiment because the boundary conditions at the top and
bottom of the cylinder must be stress free\cite{knobloch,alonso95}, a
requirement that cannot be attained in terrestrial RBC experiments.   In
contrast, two-dimensionality, stress free surface conditions and radial
driving forces all arise naturally in the electroconvection of an annular
suspended film. Thus, the
vortices that occur in annular electroconvection without shear are accurate 2D 
analogs of `Taylor columns'.

The similarity between our system and many other better-studied systems and models 
suggests that many linear and nonlinear techniques developed for other problems
can be fruitfully brought to bear on annular electroconvection.  In addition, the system
may allow the study of bifurcation scenarios that are not experimentally realizable in other similar systems.

.

\section{Conclusion}
\label{conclusion}

In this paper, we have reported a wide ranging experimental study of the primary bifurcation
to electroconvection in sheared two-dimensional annular 
films. Our principal experimental probe was the excess current carried by the film
due to convection, which can be directly related to the amplitude of the convective flow.
  In all, we examined annuli with six different radius ratios $\alpha$, with
$0.33~\leq~\alpha~\leq~0.80$.  For these, the Reynolds number
of the applied Couette shear varied between $0~\leq~{\cal R}e~<~3$.  The
explored range for the dimensionless parameter ${\cal P}$, which varies with film thickness 
and conductivity,
 was $1~<~{\cal P}~<~150$. 

We compared this data, in the first instance, with the predictions of linear theory\cite{DDM_pof_99}.
The data for the critical  voltage  $V_c^0$ for the onset of convection without shear could be
combined for all six radius ratios and was shown to obey the scaling predicted by
linear theory. This data could then be used in a slightly model-dependent way to fix the only
remaining  unknown material parameter, the fluid viscosity.  We could then compare the 
experimentally measured suppression of the onset by shear to linear theory
in an essentially pararmeter-free way.  The agreement was satisfactory for a wide range of 
$\alpha$, ${\cal P}$, and ${\cal R}e$.  Using a simple visualization scheme, we also confirmed 
that the azimuthal mode number $m$ near onset was close to the most unstable critical value $m_c$ 
predicted by linear theory.

Nonlinear fits to current-voltage data above the onset of convection were then used to infer
the real part of the amplitude of convection.  Under shear, the amplitude can be shown to obey
a complex Ginzburg-Landau equation; its real part was fit to the time-independent steady state
version of this equation in order to extract the coefficient $g$ of the cubic nonlinearity.
When shear is absent, the primary bifurcation is a pitchfork while with shear, the bifurcation
is a pitchfork Hopf.  We determined the functional dependence of $g$ as $\alpha$, ${\cal R}e$
and ${\cal P}$ were varied.  When shear is absent, it was found that for $\alpha =
0.47,0.56,0.60,0.64,0.80$ and ${\cal P} \stackrel{>}{\sim} 13$ that $g$ was independent of
${\cal P}$.  In addition, in this regime $g~>~0$, hence the bifurcation was supercritical.  
For $\alpha~=~0.33$, $g$ was found to be an increasing function of ${\cal P}$ for $2~<~{\cal
P}~<8$.  More importantly, it was found that the bifurcation was subcritical($g~<0$) for
${\cal P} \stackrel{<}{\sim} 5$ and supercritical($g~>0$) for ${\cal P} \stackrel{>}{\sim} 5$.  
In overall trend, we found that $g$ is an increasing function of $\alpha$ for zero shear.  
The largest $\alpha = 0.80$ was sufficiently close to the limiting case of $\alpha \rightarrow
1$ that we could extrapolate the measured values of $g$ for comparison to the value of $g$
calculated from weakly nonlinear theory for the corresponding `plate' electrode geometry.  
This quantitative comparison gave reasonable agreement.

When shear was applied, it had a strong effect on the subcriticality of the primary
bifurcation. Measurements of $g$ as a function of the shear Reynolds number ${\cal R}e$
revealed that for $\alpha = 0.47,0.56,0.60,0.64,0.80$ there existed a tricritical Reynolds
number ${\cal R}e_T$ below which $g > 0$ and the onset is a supercritical Hopf bifurcation.
For ${\cal R}e > {\cal R}e_T$, $g < 0$ and the onset became a subcritical Hopf bifurcation
with the degree of subcriticality a nontrivial function of ${\cal R}e$.  Hence, the Reynolds
number forms an interesting second control parameter in this system which can be used to vary
the nature of the primary bifurcation.

The data presented in this paper represents only a sparse sampling of the full 3D parameter
space of $g(\alpha, {\cal P}, {\cal R}e)$. Within this space, there presumably exist
continuous 2D loci on which $g=0$ and the primary bifurcation is tricritical.  On other loci,
the hysteresis $\delta\epsilon$ is a local maximum. As a function of $\alpha$, linear
theory\cite{linear} shows that there exist special values of $\alpha$ at which two azimuthal
mode numbers $m$ and $m+1$ are simultaneously linearly unstable. All of the $g$ results we
have described pertain to the {\it primary} bifurcation; above onset we also observed numerous
{\it secondary} bifurcations which take the form of hysteretic jumps in $m$.  The location of
these secondary bifurcations is a function of ${\cal R}e$.
  These considerations suggest that more-or-less discontinuous jumps in the behavior of $g$
may occur at parameter values across which the azimuthal mode number $m$ of the {\it
nonlinear} pattern changes discontinuously. The rich phenomenology of even the primary
bifurcation presents a significant challenge to the weakly nonlinear theory of this
instability.

\section*{acknowledgements}
This work was supported by the Natural  Science and Engineering Research Council of Canada and the 
U.S. Department of Energy under contract W-7405-ENG-36.

\vskip 1in

\begin{table}
\center
\begin{tabular}{|c|c|c|}\hline
radius ratio&Experimental&Theoretical\\
$\alpha$&$m_c^0$&$m_c^0$\\ \hline
0.33&4&4\\
0.47&6&6 \\
0.56&8&7 \\
0.60&8&8 \\
0.64&10&10 \\
0.80&20&19 \\ \hline
\end{tabular}

\caption{Experimental measurements of the marginally stable mode number for
zero shear,
$m_c^0$.Theoretical values are from Ref.~[3]. }
\label{marginalmode}
\end{table}
\vskip 0.2in

\begin{table}[t]
\center
\begin{tabular}{|c|c|c|c|}\hline
radius ratio&Experimental&${\cal P}$&Theoretical\\
$\alpha$&$g$&range&$g$\\ \hline
0.33&-0.74 $\pm$ 0.23&$2.1~<~{\cal P}~<~4.4$& \\
0.47&~1.64 $\pm$ 0.06&$13.5~<~{\cal P}~<~20.7$& \\
0.56&~0.73 $\pm$ 0.15&$59.4~<~{\cal P}~<~100.8$& \\
0.60&~2.72 $\pm$ 0.34&$31.3~<~{\cal P}~<~38.9$& \\
0.64&~1.87 $\pm$ 0.10&$25.2~<~{\cal P}~<~63.0$& \\
0.80&~2.21 $\pm$ 0.29&$15.3~<~{\cal P}~<~142.8$& \\  \hline
1.00~(`plate')&~&${\cal P} = \infty$&2.842 \\  \hline
\end{tabular}
\vskip 0.15in
\caption{Experimental measurements of the
coefficient of the cubic nonlinearity, $g$ without shear.}
\label{gvsrrtable}
\end{table}
\vskip 0.2in

\begin{table}
\center
\begin{tabular}{|c|c|c|}\hline
radius ratio&Reynolds number&${\cal P}$\\
$\alpha$&${\cal R}e_T$&range\\ \hline
0.47&0.18~$\pm$~0.02&$15.8~<~{\cal P}~<~16.6$ \\
0.56&0.03~$\pm$~0.02&$75.4~<~{\cal P}~<~85.4$ \\
0.60&0.03~$\pm$~0.01&$30.3~<~{\cal P}~<~31.7$ \\
0.64&0.08~$\pm$~0.06&$29.1~<~{\cal P}~<~61.2$ \\
0.80&0.01~$\pm$~0.01&$65.5~<~{\cal P}~<~70.6$ \\ \hline
\end{tabular}
\caption{Experimental measurements of the tricritical Reynolds number ${\cal R}e_T$ at which $g = 0$.}
\label{Reg=0}
\end{table}
\vskip 0.2in

\begin{table}
\center
\begin{tabular}{|c|c|c|c|c|}\hline
radius ratio&Minimum&Reynolds number&~&Maximum \\
$\alpha$&$g$&${\cal R}e_{min}$&${\cal P}$&${\cal R}e$ \\ \hline
0.47&-3.68~$\pm$~0.19&0.83~$\pm$~0.18&15.3&2.59 \\
0.56&-5.15~$\pm$~1.04&0.11~$\pm$~0.05&63.3&0.22 \\
0.60&-1.74~$\pm$~0.04&0.05~$\pm$~0.02&32.1&0.13\\
0.64&-4.34~$\pm$~0.79&0.23~$\pm$~0.02&53.4&0.25\\
0.80&-9.17~$\pm$~0.56&0.04~$\pm$~0.01&12.0&0.10\\  \hline
\end{tabular}

\caption{Experimental measurements
of the minimum value of $g$ and the corresponding ${\cal R}e_{min}$ and ${\cal P}$
values.}
\label{ming}
\end{table}


\begin{references}
\bibitem{mc_note} It was also found that the critical azimuthal mode number
under shear is reduced so that $m_c({\cal R}e > 0) \leq {m_c}^0$.
\bibitem{dope_note} Samples with high concentrations of TCNQ, left to
themselves for a period of weeks to months, changed color from off-white to
orange to green.  The same changes were observed in a solution of TCNQ in
acetonitrile independent of concentration.  At the lower dopant
concentrations used in the experiments, the samples remained an off-white
color.
\end{references}
\end{document}